\documentclass[letterpaper,12pt]{article} 

\usepackage{amsmath}
\usepackage{amssymb}
\usepackage[round]{natbib}
\usepackage[dvips]{epsfig}
\usepackage{dcolumn}
\usepackage{enumerate}
\usepackage{hhline}
\usepackage{dsfont}
\usepackage{afterpage}
\usepackage{arydshln}
\usepackage{graphicx}
\usepackage{color}
\usepackage[usenames,dvipsnames]{xcolor}
\usepackage{rotating}
\usepackage[breaklinks]{hyperref}
\usepackage{breakurl} 
\usepackage{xr}
\usepackage[percent]{overpic}
\usepackage{subfig}
\usepackage[]{caption}

\usepackage{algorithmicx}
\usepackage[noend]{algpseudocode}
\usepackage{algorithm}
\usepackage{diagbox}
\usepackage{graphicx}
\usepackage{wrapfig}
\usepackage{lscape}
\input epsf
\usepackage{fontenc}
\usepackage{setspace}
\usepackage{bm}
\usepackage{slashbox}
\usepackage{lscape}
\usepackage{breakurl} 
\usepackage{multirow}
\usepackage{eurosym}
\epsfverbosetrue
\def\theequation{\thesection.\arabic{equation}}  
\def\abstract{\if@twocolumn
\section*{Abstract}
\else \normalsize 
\begin{center}
{\bf Abstract\vspace{-.5em}\vspace{0pt}} 
\end{center}
\quotation 
\fi}
\def\endabstract{\if@twocolumn\else\endquotation\fi}

\makeatletter
\newcommand{\myappendix}[1]{
	\setcounter{section}{1}
        \renewcommand{\thesection}{A\arabic{section}}}

\def \dsR {\text{$\mathds{R}$}}

\DeclareMathOperator{\diag}{diag}

\DeclareMathOperator{\ND}{N}

\def \dvec {\text{\boldmath$d$}}

\def \uvec {\text{\boldmath$u$}}

\def \xvec {\text{\boldmath$x$}}    
\def \yvec {\text{\boldmath$y$}}    
\def \zvec {\text{\boldmath$z$}}

\def \betavec         {\text{\boldmath$\beta$}}

\def \varepsilonvec   {\text{\boldmath$\varepsilon$}}
\def \zetavec         {\text{\boldmath$\zeta$}}
\def \etavec          {\text{\boldmath$\eta$}}
\def \thetavec        {\text{\boldmath$\theta$}}

\def \lambdavec       {\text{\boldmath$\lambda$}}

\def \rhovec          {\text{\boldmath$\rho$}}

\def \psivec          {\text{\boldmath$\psi$}}

\usepackage{color}
\usepackage{colordvi}
\newlength{\breite}
\breite\textwidth
\newcounter{aufg}[section]
  {\refstepcounter{aufg}\noindent\textbf{Exercise \arabic{aufg}:}
   \\*[1ex]\noindent}{\vspace{.5cm}}
   
 \newcounter{notes}[section]
  {\refstepcounter{aufg}\noindent\textbf{}
   \\*[1ex]\noindent}{\vspace{.5cm}}
   
\usepackage{amsthm}  

\theoremstyle{definition}

\newtheorem*{beisp*}{Example}
\newtheorem{Proof}{Proof}


\newtheoremstyle{break}
  {}
  {}
  {}
  {}
  {\bfseries}
  {.}
  {\newline}
  {}
  
\theoremstyle{break}

\newcommand{\head}[2]%
 {\hrule \vspace{.15cm} {\sfbold Advanced Statistical Inference, Summer Term 2012, Georg-August-University G\"ottingen}\hfill
{\sfbold Sheet #1}\\
{\sfbold Prof. Dr. Thomas Kneib, Nadja Klein}\hfill {\sfbold #2}

\vspace{.2cm}
\hrule

\vspace{1cm}

}

\newcounter{auf}
{\refstepcounter{auf}
\begin{center}
\fcolorbox[gray]{0}{.95}{
\makebox[\breite]{
\textbf{Exercise \arabic{auf}}
}}\\*[1ex]\noindent
\end{center}
}{\vspace{.5cm}}

\newcounter{loes}[section]
{\stepcounter{loes}
\begin{center}
\fcolorbox[gray]{0}{.95}{
\makebox[\breite]{
\textbf{L"osung \arabic{loes}}
}}\\*[1ex]\noindent
\end{center}
}{}

{\begin{center}
\fcolorbox[gray]{0}{.95}{
\makebox[\breite]{
\textbf{Zu Aufgabe #1}
}}\\*[1ex]\noindent
\end{center}\vspace{1cm}
}{\vspace{1cm}}

\newcounter{ka}
{\refstepcounter{ka}
\begin{center}
\framebox[\textwidth]{
\textbf{Aufgabe \arabic{ka}} \hfill #1 Punkte
}\\*[1ex]\noindent
\end{center}
}{\vspace{1cm}}

\newcounter{lka}
{\refstepcounter{lka}
\begin{center}
\framebox[\textwidth]{
\textbf{L\"osung \arabic{lka}} \hfill #1 Punkte
}\\*[1ex]\noindent
\end{center}
}{\vspace{1cm}}

\renewcommand{\baselinestretch}{1.5}

\newcounter{myremark}

\newcounter{mynotation}

\usepackage{paralist}

\renewenvironment{itemize}[1]{\begin{compactitem}#1}{\end{compactitem}}

\makeatletter
\def\@seccntformat#1{\@ifundefined{#1@cntformat}%
	{\csname the#1\endcsname\quad}  
	{\csname #1@cntformat\endcsname}
}
\let\oldappendix\appendix 
\renewcommand\appendix{%
	\oldappendix
	\newcommand{\section@cntformat}{\appendixname~\thesection\quad}
}
\makeatother

\usepackage{titlesec}

\titlespacing*\section{0pt}{0pt plus 4pt minus 2pt}{0pt plus 2pt minus 2pt}
\titlespacing*\subsection{0pt}{0pt plus 4pt minus 2pt}{0pt plus 2pt minus 2pt}
\titlespacing*\subsubsection{0pt}{0pt plus 4pt minus 2pt}{0pt plus 2pt minus 2pt}
\titlespacing*\paragraph{0pt}{5pt plus 4pt minus 2pt}{7pt plus 2pt minus 2pt}

\usepackage{scalerel,stackengine}
\stackMath
\newcommand\reallywidehat[1]{%
\savestack{\tmpbox}{\stretchto{%
  \scaleto{%
    \scalerel*[\widthof{\ensuremath{#1}}]{\kern-.6pt\bigwedge\kern-.6pt}%
    {\rule[-\textheight/2]{1ex}{\textheight}}
  }{\textheight}%
}{0.5ex}}%
\stackon[1pt]{#1}{\tmpbox}%
}

\begin{document}
\pagestyle{empty}

\title{Marginally-calibrated deep distributional regression}

\author{Nadja Klein\footnote{Nadja Klein is Assistant Professor of Applied Statistics at Humboldt-Universit\"at zu Berlin}\,\,\footnote{Communicating Author: {\tt nadja.klein@hu-berlin.de}.  Routines required to estimate the DNNC for Section 4 are provided as part of the supplementary material.}\,\,, David J. Nott\footnote{David J. Nott is Associate Professor of Statistics and Applied Probability at National University of Singapore}\,\, and Michael Stanley Smith\footnote{Michael Stanley Smith is Professor of Management (Econometrics) at Melbourne Business School, University of Melbourne.}} 
\date{}
\maketitle
\noindent

\begin{abstract}
\noindent
Deep neural network (DNN) regression models are widely used in applications requiring state-of-the-art predictive accuracy. However, until recently 
there has been little work on
accurate uncertainty quantification for predictions from such models. We add to this literature by outlining 
an approach to constructing predictive distributions 
that are `marginally calibrated'. This is where the long run average 
of the predictive distributions of the response variable matches the observed empirical
margin. 
Our approach considers a DNN regression with a conditionally Gaussian prior for the final layer weights, from which an implicit copula process 
on the feature space is extracted. 
This copula process is combined with a non-parametrically estimated 
marginal distribution for the response. The end result is 
a scalable distributional DNN regression method with marginally calibrated predictions,
and our work complements
existing methods for probability calibration.
The approach is first illustrated using two applications of 
dense layer feed-forward neural networks. 
However, our main motivating applications are in likelihood-free inference, where distributional deep regression is used to estimate marginal posterior distributions.
In two complex ecological time series examples we employ
the implicit copulas of convolutional networks, and show that
 marginal calibration results in improved uncertainty quantification.
 Our approach also avoids the need for manual specification of summary statistics, 
 a requirement that is burdensome for users and typical of competing likelihood-free inference methods.
 \vspace{20pt}
 
\noindent
{\bf Keywords}: Calibration, Copula, Deep Neural Network, Distributional Regression, Likelihood-free Inference, Uncertainty Quantification.
\end{abstract}


\newpage
\pagestyle{plain}
\setcounter{page}{1}
\setcounter{equation}{0}
\renewcommand{\theequation}{\arabic{equation}}
\setlength{\abovedisplayskip}{0.1cm}
\setlength{\belowdisplayskip}{0.1cm}

\section{Introduction}
Deep models have become very popular in 
applications requiring high predictive accuracy \citep{goodfellow+bc16}.
In addition to being flexible, they
are scalable to large datasets with high-dimensional features.  
However, in some applications it
is crucial to represent 
uncertainty in predictions accurately, which is 
something that na\"{i}ve applications of deep learning models often fail to do.  
This has motivated recent research on extensions that are able to capture aspects of the predictive
distributions beyond the mean, and on methods to calibrate
the full predictive distributions accurately.
Our aim in the present work is 
to review and add
to this existing literature in the area of distributional
calibration methods for deep neural network (DNN) regression. 

To do so we 
develop a new scalable method for `distributional deep regression', by which 
we mean a DNN regression method that provides predictions for the full distribution. 
The proposed method uses
the implicit copula~\citep[p.51]{nelsen06} of a vector of values on a response variable
that arises from a DNN regression. 
We call this variable a `pseudo-response' because it is not observed directly. 
The resulting copula is a highly flexible deep function of the feature vector, 
which we combine with a non-parametrically estimated marginal distribution for the 
observed response variable. The predictive distributions from the model
are marginally-calibrated (where the long-run average of the predictive distributions 
matches the empirically observed margin) 
and the approach extends the marginally-calibrated 
regression copula models of \citet{klein+s18} and \cite{smith+k19}
to deep models.
Even though the proposed copula is of very high dimension, we show that the likelihood is easy to compute 
using  Bayesian methods and existing neural net libraries optimized for scalability.

Importantly,
all aspects of the predictive distribution---such as the mean, variance, higher order moments
and tail behaviour---are learned jointly in this deep regression.
 To illustrate this, and other
advantages of our approach, we first consider the implicit copula of a dense
layer feed-forward network, and apply the resulting distributional deep regression to
two popular benchmark
datasets. In both cases, our approach provides substantially more accurate
predictive densities,
compared to those obtained from applying 
the feed-forward network directly to the data with, or without, probability calibration.

However, our main application of this new distributional deep regression model
is in likelihood-free inference. Here,
we estimate Bayesian posterior distributions for models with intractable likelihood functions.
In this case, distributional regression methods can be employed 
to estimate the posterior density using 
training samples simulated from the joint model, with 
the parameter as the response variable and the data as feature values. After fitting the distributional regression, the approximate posterior is obtained as
the predictive distribution with feature values given by the observed data.
Our approach is quite general,
but in likelihood-free inference applications we show that the marginal calibration property
acts to improve uncertainty quantification significantly.  

To illustrate the advantages of our approach in likelihood-free inference,
we construct the implicit copula of a convolutional network, and apply the resulting
distributional deep regression to compute inference for two complex
applications in ecological time series considered in \citet{wood10} and \citet{FasWoo2016}.   
\citet{wood10} considered the use of likelihood-free inference methods based on summary statistics for inference in state space models
where the likelihood may be highly irregular.  
He developed an approximate likelihood, called the synthetic likelihood, which is based on a Gaussian model for a vector-valued 
non-sufficient summary statistic, where
the summary statistic mean and covariance are estimated for each parameter value by Monte Carlo simulation.  
Discarding some information by using a non-sufficient summary of the data can lead to a better behaved likelihood function.  
A recent comparison of full likelihood and synthetic likelihood inference
is given in \citet{FasWoo2016}, where some typical applications in ecology and epidemiology are described.
Bayesian implementations of synthetic likelihood are discussed in detail in \citet{price+dln16}.
   
There are alternative likelihood-free inference methods for time series data, most notably the approximate Bayesian computation (ABC)
framework \citep{sisson+fb18}. A recent discussion of ABC methods for time series is given in \citet{frazier+mmm19} who 
suggest that accurate estimation
of the posterior on parameters is not always necessary for 
accurate forecasting.  
However, like synthetic likelihood and its extensions, ABC methods require suitable summary statistics for the data.  
Even without the requirement that these statistics be multivariate normal for every parameter value, 
choosing statistics that are informative and low-dimensional is difficult. 
A major advantage of our proposed approach is that  
manually specified summary statistics are not required,
and instead the whole dataset is used as the feature vector in a
regression model. 
A further comparison of our approach and alternative likelihood-free inference methods
is given in Section~\ref{sec:LFI}.

The rest of the paper is organized as follows. Section~\ref{sec:dlreg} 
introduces deep learning regression models, and 
uncertainty quantification of predictions from such models. Section~\ref{sec:DNN} shows how the 
copula smoother of \citet{klein+s18} can be extended to the DNN case, including computation of the predictive distributions.  Section~\ref{sec:FNN} 
illustrates our approach using dense layer feed-forward networks for two benchmark datasets. Section~\ref{sec:LFI} reviews likelihood-free inference, and 
uses an implicit copula constructed from a convolutional network to perform 
likelihood-free inference in two complex ecological time series models. A comparison to
some leading alternative approaches is also provided, while Section 6 concludes.  

\setlength{\abovedisplayskip}{0.1cm}
\setlength{\belowdisplayskip}{0.1cm}

\section{Deep Learning for Regression}\label{sec:dlreg}
In this section we briefly introduce deep learning regression models, and review the existing
literature on uncertainty quantification for deep learning. 

\subsection{Deep learning regression models}
Suppose we observe $n$  
response and feature values $(Z_i,\xvec_i)$, $i=1,\dots, n$, where $Z_i$ is the 
scalar
response and $\xvec_i=(x_{i1},\dots, x_{ip})^\top$ the vector of 
$p$ feature values.
A feed-forward neural network 
specifies a function of the input features through 
a sequence of transformations
called `layers'. The feature vector, $\xvec_i$, enters the model at an initial input layer,
while the final transformation gives the predicted response at an output layer; the
 intermediate component transformations are the hidden layers.  The inputs to any layer
come from previous layers.  
For an introduction to neural network terminology from a statistical perspective
see overviews by~\cite{polson+s17} and~\cite{fan2019}.

In this paper we consider DNNs for regression with a single response.
A DNN can be written as a function $f_\etavec(\xvec_i)$,
 where $\etavec$ is the set of all parameters 
(weights) in the network. 
We assume that the activation function for the output layer is linear, so that 
\begin{align}
  f_\etavec(\xvec_i) & = \psi_{\zetavec}(\xvec_i)^\top \betavec+\beta_0\,,    \label{network}
\end{align}
where $\betavec$ are the output layer coefficients, $\zetavec$ are the coefficients 
of all other layers (so that $\etavec=(\beta_0,\betavec^\top,\zetavec^\top)^\top$), and $\psi_\zetavec(\cdot)$
is the vector of $q$ basis functions defined by the last hidden layer of the network.
In general an intercept $\beta_0$ is also included, although in Section~\ref{sec:DNN}
we set $\beta_0=0$ in the output layer 
because implicit copulas are location free.\footnote{Note
that in Section~\ref{sec:DNN} the random variable $Z_i$ is a pseudo-response, 
and we construct the implicit copula
of its data distribution.} 

Training of a neural network such as (\ref{network}) is usually done by minimizing a penalized empirical loss function with respect
to $\etavec$. A popular loss in regression problems is the squared error over the
observed values $\zvec=(z_1,\ldots,z_n)^\top$ of the response, given by
\begin{align}
  L(\bm{z},f_\etavec) & = \sum_{i=1}^n (z_i-f_\etavec(\xvec_i))^2\,. \label{l2loss}
\end{align}
A regularization penalty---such as the  $L2$ or $L1$ norm of the weights---is often used to prevent over-fitting and simplify the optimization. 
 Regularization can also be undertaken in an implicit manner in the optimization 
 algorithm, using a variety of methods such as early stopping, drop-out or batch normalization; see
\citet{goodfellow+bc16} for an introduction to these ideas.
For posterior mode estimation in a Bayesian framework, the loss is equivalent to the negative log-likelihood and an explicit penalty is 
equivalent to the negative log prior density. 
When (\ref{l2loss}) is used in a regression setting, it is equivalent to assuming a homoscedastic Gaussian model, although minimization of (2) can be justified on grounds which do not require parametric model assumptions.
Thus, the simplest way to construct predictive distributions for future responses is to use a Gaussian predictive density, 
after estimating the response variance. However, in many applications  
more sophisticated uncertainty quantification is necessary, which is the focus of the current paper.

In our approach we  consider an initial DNN regression fit, from which
data-dependent basis functions $\psi_\zetavec(\cdot)$ are obtained.
These basis functions are constructed using optimized standard neural network libraries, which are then employed in our Bayesian copula regression model outlined in Section~\ref{sec:DNN}. 

Our methodology is related to that of \citet{nalenz+v18}, where
data-dependent basis functions from tree ensemble methods were further used within a Bayesian statistical model with shrinkage priors
to obtain uncertainty quantification. It is also related to
a `neural linear model' \citep{ober2019}, where a statistical 
linear model is used for the output layer of a DNN regression, because we consider a Bayesian neural linear model for a pseudo-response
in Section~\ref{sec:regcop}.

\subsection{Uncertainty quantification for deep learning}\label{sec:uncertDNN}
\cite{gneiting+br07} discuss different ways to gauge the accuracy of uncertainty 
quantification in prediction, and we focus on two of these: `probability calibration' and `marginal calibration'. 
\citet{gneiting+br07} define notions of calibration formally by considering a 
sequence of forecasting problems indexed by time.  
``Nature" chooses a distribution $H_t$ at time $t$, from which an outcome is drawn, and the forecaster gives a predictive
distribution $F_t$.  Intuitively, 
probability calibration means that an event 
predicted to have probability $p$  
occurs with relative frequency $p$, formalized by the requirement that
\[
\lim_{T\rightarrow\infty} \frac{1}{T}\sum_{t=1}^T H_t\circ F_t^{-1}(p)=p\,,\;\;\mbox{ for all } p\in(0,1)\,.
\]
Roughly speaking, 
marginal calibration is where the average of nature's true distribution should match
the average forecast distribution, so that
\begin{align*}
\lim_{T\rightarrow\infty} \frac{1}{T}\sum_{t=1}^T H_t & = \lim_{T\rightarrow\infty} \frac{1}{T} \sum_{t=1}^T F_t,
\end{align*}
if the limits on the left- and right-hand sides above exist. 
In practice, while $H_t$ is unknown, a draw from it is observed and
$H_t$ can be replaced with a point mass at this value, so 
that $\frac{1}{T}\sum_{t=1}^T H_t$ is
the empirical distribution function.

In the current work, the forecaster is a regression model and $F_t$ is its predictive distribution. 
We do not follow the above theoretical framework strictly in our later development. For example, our regression is applied to cross-sectional data, so that time ordering does not apply.
 However, as noted in 
\cite{gneiting+br07}, the above framework can 
still provide related empirical 
notions of marginal calibration. \cite{gneiting+r13}
develop such a notion by considering the forecast
distribution as a random measure, and there is a joint distribution of this random measure and the observation. Marginal calibration can be defined as equality of the marginal distribution of the observation and the expected forecast distribution.  

Discussions of uncertainty quantification often make a distinction between aleatoric and epistemic uncertainty.
Aleatoric uncertainties are those which cannot be reduced
in principle, whereas epistemic uncertainties arise from knowledge that it is possible to possess, but is not
available. Examples of epistemic uncertainty
in statistical modelling are model uncertainty and parameter uncertainty.  
The distinction between aleatoric and epistemic uncertainty is important in some applications of
deep learning models, such as extrapolating predictions to regions of the 
feature space not covered by the training set  
\citep{kendall+g17}.

In the deep learning literature previous 
approaches to achieving accurate uncertainty quantification are of two main types, 
which we discuss separately below.

\subsubsection{Post-processing calibration adjustments}
The first approach  uses 
post-processing adjustments to achieve probability calibration. 
The early machine learning literature on this topic is concerned with classification problems (e.g. \citet{platt99}), 
while calibration in regression has received far less attention in the machine learning literature.  One approach for regression is due to \citet{keren+cs18}, 
who consider discretizing a continuous response so that calibration methods developed for
classifiers can be applied. 
A related method is described in \citet{li+br19}, who consider classification methods 
which account for the ordering of a finely discretized response.

Another approach, which we compare to our own later, is due to \citet{kuleshov+fe18}   
and can be briefly explained as follows. 
Suppose we have a training set of features and responses, and a regression model
has been fitted to them.  The fitted model provides predictive distributions for the response for any value of the feature vector.
Now consider a calibration set of features and responses, which may be disjoint from the training set.    
For every probability $p\in [0,1]$ we can ask what is the value $p'=p'(p)$ such that the $p'$-quantiles of the predictive 
distributions upper bound a relative frequency of $p$ of the responses in the calibration set?  
We then adjust all predictive distributions such that the $p$-quantile is changed to the corresponding $p'$-quantile.  This adjustment 
can be achieved using isotonic regression for a set of $p$ and $p'$ pairs, and 
is a form of probability calibration.
  
Our method described in Section~\ref{sec:DNN}
ensures that the marginal distribution of the model response
matches an empirical estimate. However, whether the ergodic averaging of predictive distributions leads to equality with
the model marginal distribution depends on the properties of the regression copula. However, we show that the average predictive distribution from our method reproduces the empirically estimated marginal well in the examples.  

\citet{kuleshov+fe18}, 
in their discussion of the notion of marginal calibration in \citet{gneiting+br07}, state that ``We found that their notion of
marginal calibration was too weak for our purposes, since
it only preserves guarantees relative to the average distribution.''  In contrast, we show that marginal calibration can be highly constraining at times, so that it complements probability
calibration post-processing. In our experience, it is particularly effective when the response distribution
is skewed, heavy-tailed or bounded. 

\subsubsection{Distributional regression approaches}
A second approach to uncertainty quantification for
deep learning models in regression 
is to use more flexible models that can capture apects
of the response distribution beyond the mean. For example,
loss functions equivalent to 
Gaussian log-likelihoods with heteroscedasticity have been considered by \citet{kendall+g17} and \cite{lakshminarayanan+pb17}, among others, where both the mean and log-variance
are flexible functions of the features. 
However, with such an approach the response distribution 
is still conditionally Gaussian, which will result in a lack of calibration in some problems.  These authors also consider
various other innovations designed to improve uncertainty quantification, including uncertainties relating to feature
vectors which are unusual compared to the training set.  
A deep version of quantile regression
has been recently considered by \citet{tagasovska+l18} as a method for
modelling the whole distribution, but enforcing monotonicity
in the estimated quantiles is difficult. \citet{rodriguez+p18} consider a deep multi-task quantile learning approach which can help to avoid
the crossing quantiles problem.
Mixture density networks and their extensions are another neural approach to distributional regression possessing 
a universal approximation property
\citep{bishop94,uria+ml13}.  \citet{tran+nnk18} consider using deep learning predictors within generalized linear and mixed models, as do \citet{hubin+sf18} who also focus on the difficult problem of accounting for model uncertainty in the architecture and suggesting Markov chain Monte Carlo (MCMC) algorithms for computation.

Beyond neural network methods, there are a wide variety of 
methods for distributional regression in the statistical and machine learning literatures.  Approaches include 
Bayesian non-parametric methods
\citep{foti+w15} and the generalized additive models for location, scale and shape
framework of \citet{rigby+s05}.  The latter authors suggest using parametric response distributions beyond the exponential family, and model the mean, scale and shape parameters
as functions of the covariates;
see also \citet{mayr+fhks11,klein+kl15} and \citet{umlauf+kz18}.  A problem with many existing distributional regression methods is that they do not scale well to
large datasets, and computationally cheaper distributional regression methods are necessary in some applications.

Bayesian methods for training neural networks are also motivated by the need for improved uncertainty
quantification.  
Bayesian predictive distributions, where parameter 
and possibly model uncertainty is integrated out 
according to the posterior distribution, is a convenient way to account for these epistemic uncertainties in prediction.  
Recent work in this direction includes \citet{blundell+ckw15,hernandez-lobato+a15,gal+g16,khan+ntlgs18,kingma+sw15} and
\citet{teye+as18} among others, while
\citet{mackay92} and \citet{neal96} were pioneers of Bayesian neural
networks. A recent review of deep learning methods emphasizing the connections between existing algorithms and models
and Bayesian inference is given by \citet{polson+s17}.  
\setlength{\abovedisplayskip}{0.1cm}
\setlength{\belowdisplayskip}{0.1cm}
\section{Marginally-Calibrated Deep Learning Regression}\label{sec:DNN}
\cite{klein+s18} and~\cite{smith+k19} introduce a new approach to distributional
regression that uses a copula decomposition to ensure marginal calibration. 
We outline how to extend their approach to deep learning regression. 
\subsection{Copula model}
Consider $n$ realizations $\bm{Y}=(Y_1,\ldots,Y_n)^\top$ of a continuous-valued response,
with corresponding feature values $\xvec=\{\xvec_1,\ldots,\xvec_n\}$. Following~\cite{sklar59},
the joint density of the distribution $\bm{Y}|\xvec$ can always be written as
\begin{equation}
p(\yvec|\xvec)=c^\dagger(F(y_1|\xvec_1),\ldots,F(y_n|\xvec_n)|\xvec)\prod_{i=1}^n p(y_i|\xvec_i)\,,\label{eq:cmod1}
\end{equation}
with $\yvec=(y_1,\ldots,y_n)^\top$.
Here, $c^\dagger(\uvec|\xvec)$ is a $n$-dimensional copula density with $\uvec=(u_1,\dots,u_n)^\top$,
and $F(y_i|\xvec_i)$ is the distribution function of $Y_i|\xvec_i$; both of which are typically unknown. \cite{smith+k19} call such a copula a 
`regression copula'\footnote{This should not be confused with the term
	`copula regression' which is sometimes used to refer to a low-dimensional
	copula model for a multivariate
	response with regression margins.} 
because it is a function
of the features $\xvec$.
In copula modelling it is common
to replace $c^\dagger$ in~\eqref{eq:cmod1} with the density of a parametric copula with parameters $\thetavec$, 
and we do so here with  $c_{\mbox{\tiny DNN}}(\uvec|\xvec,\thetavec)$, which 
is the implicit copula of a DNN regression specified below
in Section~\ref{sec:regcop}. 
 \cite{klein+s18} suggest calibrating the distribution of $Y_i|\xvec_i$ to 
 its invariant margin, so that density $p(y_i|\xvec_i)
 =p_Y(y_i)$ with distribution function $F_Y$ estimated non-parametrically.
Thus, the copula model is
\begin{equation}
p(\yvec|\xvec,\thetavec)=c_{\mbox{\tiny DNN}}(F_Y(y_1),\ldots,F_Y(y_n)|\xvec,\thetavec)\prod_{i=1}^n p_Y(y_i)\,.\label{eq:cmod2}
\end{equation}
We stress that 
even though $Y_i|\xvec_i$ is assumed invariant with respect to $\xvec_i$, the response
is still affected by the features $\xvec$ though the joint distribution, which~\cite{smith+k19} point out has two consequences. First, the entire marginal predictive distribution
of a future response $Y_{0}$ given in Section~\ref{sec:pdens} is a function of the feature vector $\xvec_{0}$.
Second, the implicit copula developed below has some additional latent parameters $\betavec$, and the
marginal distribution $Y_i|\xvec_i,\betavec,\thetavec$ that also 
conditions on $\betavec$ is also a function of $\xvec_i$, analogous to the
specification of a standard regression model.

\subsection{Regression copula}\label{sec:regcop}
The key to the success of our approach is the specification of the regression copula with 
density $c_{\mbox{\tiny DNN}}$. For this we employ 
the implicit copula of a pseudo-response
vector from a DNN regression model derived as follows. Consider
a pseudo-response given by the output layer 
observed with Gaussian noise, so that if $\varepsilon_i$ is distributed independently
$N(0,\sigma^2)$, 
\begin{equation}
\tilde Z_i=f_\etavec(\xvec_i)+\varepsilon_i\,.\label{eq:rtilde}
\end{equation}
Then from~\eqref{network}, the vector of $n$ 
realizations $\tilde{\bm{Z}}=(\tilde{Z}_1,\ldots,\tilde{Z}_n)^\top$ 
is given by the linear model
\begin{equation}
\tilde{\bm{Z}}=B_\zetavec(\xvec)\betavec+\varepsilonvec\,,\;\varepsilonvec\sim N(0,\sigma^2I)\,,
\label{eq:linmod}
\end{equation}
with $B_\zetavec(\xvec)=[\psi_{\zetavec}(\xvec_1)|\cdots| \psi_{\zetavec}(\xvec_n)]^\top$ an $(n\times q)$ matrix. 
The regression \eqref{eq:linmod} does not contain an intercept, since an intercept term is not identified in the copula.
We assume that $\zetavec$ is known so that the basis functions are fixed, and the procedure for obtaining $\zetavec$ is described later.  
To produce
smooth and efficient estimates we regularize the basis coefficient vectors $\betavec$.
In a Bayesian context this corresponds to adopting a shrinkage prior, with a common 
choice being
the conditionally Gaussian prior
\[
\betavec|\thetavec,\sigma^2\sim N(\bm{0},\sigma^2 P(\thetavec)^{-1})\,,
\]
where $P(\thetavec)$ is a sparse precision matrix that is a function of regularization
parameters $\thetavec$.

We extract the copula of the distribution of the pseudo-response vector $\tilde{\bm Z}$ with
$\betavec$ integrated out. 
Such a copula is either called an `implicit'~\citep[p.190]{McNFreEmb2005} or `inversion'~\citep[p.51]{nelsen06} copula
because it is constructed by inverting Sklar's theorem. The copula is $n$-dimensional
with a dependence structure that is a function of the feature values 
$\xvec$. To derive this copula, first notice that under the linear model and Gaussian prior
\[
\tilde{\bm{Z}}|\xvec,\sigma^2,\thetavec \sim
 N\left(0,\sigma^2\left( I+B_\zetavec(\xvec)P(\thetavec)^{-1}B_\zetavec(\xvec)^\top \right) \right)\,,
\]
which is derived in Section~2.1 of~\cite{klein+s18}.
The implicit copula of this distribution is called the Gaussian copula, and it is constructed
by standardizing the distribution above to have zero mean and unit variances. To do so 
here, we set $\bm{Z}=(Z_1,\ldots,Z_n)^\top=\sigma^{-1}S(\xvec,\thetavec)\tilde{\bm{Z}}$, where 
$S(\xvec,\thetavec)=\mbox{diag}(s_1,\ldots,s_n)$ is a diagonal scaling matrix
with elements $s_i=(1+ \psi_{\zetavec}(\xvec_i)^\top P(\thetavec)^{-1}
 \psi_{\zetavec}(\xvec_i))^{-1/2}$, which ensures that
  $Z_i|\xvec,\sigma^2,\thetavec \sim N(0,1)$. The resulting Gaussian copula has density
  \begin{equation}
 c_{\mbox{\tiny DNN}}(\uvec|\xvec,\thetavec) = 
 \frac{p(\zvec|\xvec,\sigma^2,\thetavec)}{\prod_{i=1}^n p(z_i|\xvec,\sigma^2,\thetavec)}=
 \frac{\phi_n(\zvec;\bm{0},R(\xvec,\thetavec))}{\prod_{i=1}^n \phi_1(z_i)}\,,
 \label{eq:copdens}
 \end{equation}
 where 
 \begin{equation}
 R(\xvec,\thetavec) = S(\xvec,\thetavec)\left( I+B_\zetavec(\xvec)P(\thetavec)^{-1}B_\zetavec(\xvec)^\top \right) S(\xvec,\thetavec)\,,
 \label{eq:omega}
  \end{equation}
 $z_i=\Phi_1^{-1}(u_i)$, $\zvec=(z_1,\ldots,z_n)^\top$, and $\phi_n(\cdot;\bm{0},R)$ and $\phi_1$ are the densities of $N_n(\bm{0},R)$
  and $N(0,1)$ distributions, respectively. 

  Because $\sigma^2$ does not feature in the expression for $c_{\mbox{\tiny DNN}}$,
   it is therefore unidentified, so that we simply set it equal
   to 1 throughout the rest of the paper.\footnote{We stress that this does not mean the observed response $Y_i$ has unit variance, but instead has a marginal variance
   	given by $F_Y$.}
   This is because implicit copulas are always
   invariant to the scale of the pseudo-response. Also, while the copula is $n$-dimensional---and therefore potentially of very high dimension---the matrix 
   $R$ at~\eqref{eq:omega} is a parsimonious function of $\thetavec$.
   \cite{klein+s18} give expressions for $R$ for three different shrinkage priors, and we consider two different choices here:
\paragraph{Horseshoe:} The horseshoe prior is attractive due to its robustness, local adaptivity and
analytical properties~\citep{CarPol2010}.
It is a scale mixture, where $\beta_j|\lambda_j\sim N(0,\lambda_j^2)$, with
prior $\pi_0(\lambda_j|\tau)=\mbox{Half-Cauchy}(0,\tau)$ and 
$\pi_0(\tau)=\mbox{Half-Cauchy}(0,1)$.
With this prior $\thetavec=\lbrace \lambdavec,\tau\rbrace$, with
$\lambdavec=(\lambda_1,\ldots,\lambda_{q})^\top$, and 
$R(\xvec,\bm{\theta})=S(\bm{x},\bm{\theta})\left(I+B_\zetavec(\xvec)\diag(\lambdavec)^2B_\zetavec(\xvec)^\top\right)
S(\bm{x},\bm{\theta})$.
\paragraph{Ridge:} The ridge prior is one of the simplest forms of shrinkage priors, where $\beta_j|\tau^2\sim N(0,\tau^2)$ and we use the scale-dependent prior of~\cite{KleKne2016} for $\tau^2$. 
With this prior $\thetavec=\lbrace \tau^2\rbrace$, while
$R(\xvec,\bm{\theta})=S(\bm{x},\bm{\theta})\left(I+\tau^2 B_\zetavec(\xvec)B_\zetavec(\xvec)^\top\right)
S(\bm{x},\bm{\theta})$.

To link the two regression copulas in~\eqref{eq:cmod1} and~\eqref{eq:cmod2}, it is straightforward to see that 
$c^\dagger(\uvec|\xvec)=\int c_{\mbox{\tiny DNN}}(\uvec|\xvec,\thetavec)p(\thetavec)\mbox{d}\thetavec$ for some prior density $p(\thetavec)$. \cite{klein+s18} highlight
that even though $c_{\mbox{\tiny DNN}}$ is a Gaussian copula, 
$c^\dagger$ is not, although computation of $c^\dagger$
through integration with respect
to $\thetavec$ has to be undertaken numerically.
Last, the horseshoe prior has a larger copula parameter vector $\thetavec$, so that $c^\dagger$ is likely to have a richer
dependence structure. Because the horseshoe
is a `global-local' shrinkage prior, this is likely to allow for 
a more accurate dependence structure
when $\betavec$ is large and sparse as with our DNN basis functions, as we show in our later empirical work. 

\subsection{Estimation}
We employ a multi-stage estimator with the following three steps:

\noindent \underline{{\bf Algorithm~1}} {\em (Estimation of Distributional Deep Regression)}
\begin{itemize}
\item[1.] Estimate the marginal $F_Y$ using a non-parametric estimator, for 
which we use the kernel density
estimator of~\cite{shimazaki2010}. 
\item[2.] Given $F_Y$, compute pseudo-response values $z_i=\Phi^{-1}_1(F_Y(y_i))$, for $i=1,\ldots,n$. Using existing
neural net libraries applied to $\zvec=(z_1,\ldots,z_n)^\top$, 
construct output layer basis functions $\psi_\zetavec$, and evaluate them at the feature values
to obtain $B_\zetavec(\xvec)$.
\item[3.] Given $F_Y$ and $B_\zetavec(\xvec)$, compute the augmented
posterior distribution $\betavec,\thetavec|\yvec$ using MCMC, where the margin $\thetavec|\yvec$ is therefore the posterior of the copula parameters.
\end{itemize}
\vspace{10pt}

Step~2 is dependent upon the choice of architecture, which we discuss later
in the context of each application. 
Step~3 is the main challenge, where computing the posterior requires
evaluation of the likelihood, which is given by the copula decomposition at~\eqref{eq:cmod2}. To do so directly requires evaluation 
of the copula
density at~\eqref{eq:copdens}, which is computationally infeasible 
in general because of the need to invert
the $n\times n$ matrix $R$.
\cite{klein+s18} solve this problem by instead using
the likelihood conditional also on $\betavec$, which is
\begin{eqnarray*}\label{eq:clike}
p(\yvec|\xvec,\betavec,\thetavec)&= &p(\zvec|\xvec,\betavec,\thetavec)\prod_{i=1}^n 
\frac{p_{Y}(y_i)}{\phi_1(z_i)}\\
&=&\phi_n\left(\zvec;S(\xvec,\thetavec)B_\zetavec(\xvec)\betavec,
S(\xvec,\thetavec)^2\right)\prod_{i=1}^n\frac{p_{Y}(y_i)}{\phi_1(z_i)}\,,
\end{eqnarray*}
and can be evaluated in $O(n)$ operations because $S(\xvec,\thetavec)$ is diagonal. 
These authors propose standard MCMC schemes that generate $\betavec$ and $\thetavec$, so that $\betavec$
is integrated out in a Monte Carlo manner and   
direct computation of $R$ is avoided. We refer readers to~\cite{klein+s18}
for further details on these samplers, which produce J Monte Carlo draws
$\{(\betavec^{[1]},\thetavec^{[1]}),\ldots,(\betavec^{[J]},\thetavec^{[J]})\}$ from
the augmented posterior distribution $\betavec,\thetavec|\yvec$.

\subsection{Predictive densities}\label{sec:pdens}
The predictive density $p(y_0|\xvec_0)$ 
of a new observation of the response $Y_0$, given 
new feature values $\xvec_0=(x_{01},\ldots,x_{0p})^\top$, is estimated using its Bayesian 
posterior predictive density
\[
p(y_0|\xvec_0,\xvec,\bm{y}) = \int p(y_0|\xvec_0,\betavec,\thetavec)p(\betavec,\thetavec|\xvec,\bm{y})\mathrm{d}(\betavec,\thetavec).
\]
\cite{klein+s18} propose an
estimator for the above that is fast to compute, based on the Monte Carlo
draws from Step~3 of Algorithm~1. It is given by
\begin{equation}
\hat{p}_0(y_0|\xvec_0)=
\frac{p_Y(y_0)}{\phi_1(\Phi_1^{-1}(F_Y(y_0)))}
\frac{1}{\hat{s_0}}\phi_1\left( 
\frac{\Phi_1^{-1}(F_Y(y_0))-\hat{s_0}\hat{f}_\etavec(\xvec_0)}{\hat{s_0}}
\right)
\,,\label{eq:phatyx}
\end{equation}
where
 $\hat{f}_{\etavec}(\xvec_0)=\psivec_\zetavec(\xvec_0)^\top \hat\betavec$,
  $\hat{s_0}=\frac{1}{J}\sum_{j=1}^J s_0^{[j]}$,
  $s_0^{[j]}=(1+ \psi_{\zetavec}(\xvec_0)^\top P(\thetavec^{[j]})^{-1}
 \psi_{\zetavec}(\xvec_0))^{-1/2}$
and $\hat{\betavec}=\frac{1}{J}\sum_{j=1}^J \betavec^{[j]}$. Full
derivation is given in the Web Appendix.

The density forecast~\eqref{eq:phatyx} is a direct function of
the feature vector
$\xvec_0$ and is readily computed at any point in the feature space.
It is a Gaussian density for $z_0$ transformed using the nonlinear
transformation $y_0=F_Y^{-1}(\Phi(z_0))$ that does not depend on the 
features. Nevertheless, because $\hat f_\eta(\xvec_0)$ and $\hat s_0$
are highly nonlinear functions of the features via the DNN representation,
the entire density---not just the location and scale---is a flexible function of
$\xvec_0$. It is worth stressing that our approach does not have the universal approximation property of some alternative distributional regression approaches. However, the semi-parametric nature of the method allows it to be both computationally efficient, and effective with small training datasets, as we demonstrate empirically in the examples below.

\setlength{\abovedisplayskip}{0.1cm}
\setlength{\belowdisplayskip}{0.1cm}

\section{Dense Feed-Forward Network Examples}\label{sec:FNN}
To illustrate our approach we construct a regression copula from a dense layer deep feed-forward network, and apply it to two widely used benchmark regression datasets.

\paragraph{Description of datasets}
The first dataset is the Boston housing data \citep{HarRub1978} with 
$n=506$ observations and 14 features, while the second is the 
Framingham cholesterol data \citep{ZhaDav2001} with $n=1044$ observations
and 202 features.\footnote{The first dataset is available from the UCI machine learning repository, 
\url{http://archive.ics.uci.edu/ml/machine-learning-databases/housing/}, while
the second was taken from the \texttt{qrLMM} package in R. The latter is longitudinal data, where extra dummy variables
were included to account for different individuals.}
Both feature sets include binary, categorical and continuous variables,
the latter of which we standardize to the unit interval.

\paragraph{DNN architecture}
We considered our DNN copula (labelled `DNNC') with both ridge and horseshoe regularization priors. 
For the feed-forward network we used ReLU activation functions
for one to three hidden layers, along with a linear activation function for the 
output layer. 
At Step~2 of Algorithm~1, we 
obtained the basis functions of the linear activation layer using the 
package \texttt{keras} in R \citep{chollet+a18}.
We found that two hidden layers of size 64 without additional L2 regularization are sufficient  and use them combined with a dropout rate of 0.5.
In training the network early stopping was used based on ten-fold cross-validation
in both cases, with the optimization run for 200 epochs and batch size equal to the sample size. The optimizer used is adam with default settings from \texttt{keras}, see the Web Appendix for details.

\paragraph{Benchmarks}
The predictions from the DNNC are compared to four benchmarks. The first
is a feed-forward network with the same architecture above, but applied directly
to the response data $\bm{y}$ with $N(0,\sigma^2)$ disturbances and
non-zero intercept $\beta_0$ for the 
output layer (labelled `DNN'). We combine this with a ridge prior for $\betavec$.
The second benchmark is a recalibration of the predictive densities from the DNN
obtained using the approach of~\citet{kuleshov+fe18} (labelled `DNN-recalibrated').  
The third benchmark is a mixture density network implemented in the R-package \texttt{CaDENCE}~\citep{Can2012} (labelled `MDN'). We choose a two-component Gaussian mixture with all distribution parameters (the means, standard deviations and mixing weight) to be learned through a 2-hidden layer FNN.

\paragraph{Measuring accuracy}
To judge accuracy of the predictive distributions, they are evaluated  
at the observed feature values to give 
predictive densities 
$\lbrace\hat p_1(\cdot|\xvec_1),\ldots, \hat p_n(\cdot|\xvec_n)\rbrace$ using~\eqref{eq:phatyx}, from which 
distribution functions
$\lbrace\hat F_1(\cdot|\xvec_1),\ldots, \hat F_n(\cdot|\xvec_n)\rbrace$ are also computed.
From these we constructed the following three measures of accuracy.
\begin{itemize}
	\item[(i)] The first is a plot of the average predictive density 
	\[
	\hat p_{\mbox{\tiny{marg}},n}(y)\equiv\frac{1}{n}\sum_{i=1}^n \hat p_i(y|\xvec_i),
	\]
	overlaid on a histogram of the response values.
	We use this plot to assess marginal calibration of the different 
	methods. 
	\item[(ii)] The second is a plot to assess probability calibration. 
	Consider an increasing set of values $0\leq p_1<p_2<...\leq p_k\leq 1$, and define 
	\[
	\tilde p_{j}\equiv
	\frac{1}{n}\#\left\lbrace y_i\,\bigg|\,\hat F_i(y_i|\xvec_i)< p_j,\,\mbox{ }i=1,\ldots,n\right\rbrace, \mbox{ for }j=1,\ldots,k,
	\]
	to be the relative frequency of observations falling below the $p_j$-quantile of the predictive distribution, 
	where $\#$ denotes the cardinality of a set. If 
	a method produces probability calibrated predictive distributions then $\tilde{p}_j\approx p_j$. In our later examples we plot
	 $\tilde{p}_j-p_j$ versus $p_j$, with deviations from zero in the former indicating a lack of calibration. 
	\item[(iii)] The third is the mean in- and out-of-sample 
	logarithmic score. To compute the latter 
	we use ten-fold cross-validation as follows. Partition the data into ten approximately equally-sized sub-samples of sizes $n_k$,
	denoted here as
	$\{(y_{i,k},\xvec_{i,k});i=1,\ldots,n_k\}$ for $k=1,\ldots,10$.
	For each observation 
	in sub-sample $k$ we compute the predictive
	density using the remaining nine sub-samples as the training data, 
	and denote these densities here
	as $\hat p_{i,k}(y_{i,k}|\xvec_{i,k})$. 
	The ten-fold mean out-of-sample logarithmic score is then
	\[
	\mbox{MLS}=\frac{1}{10}\sum_{k=1}^{10}\frac{1}{n_k}\sum_{i=1}^{n_k}\log \hat{p}_{i,k}(y_{i,k}|\xvec_{i,k})\,.
	\]
\end{itemize}

\begin{figure}[ht]
	\caption{Predictive densities for the Boston housing data.}
	\begin{center}
\centering\includegraphics[width=0.322\textwidth,angle=0]{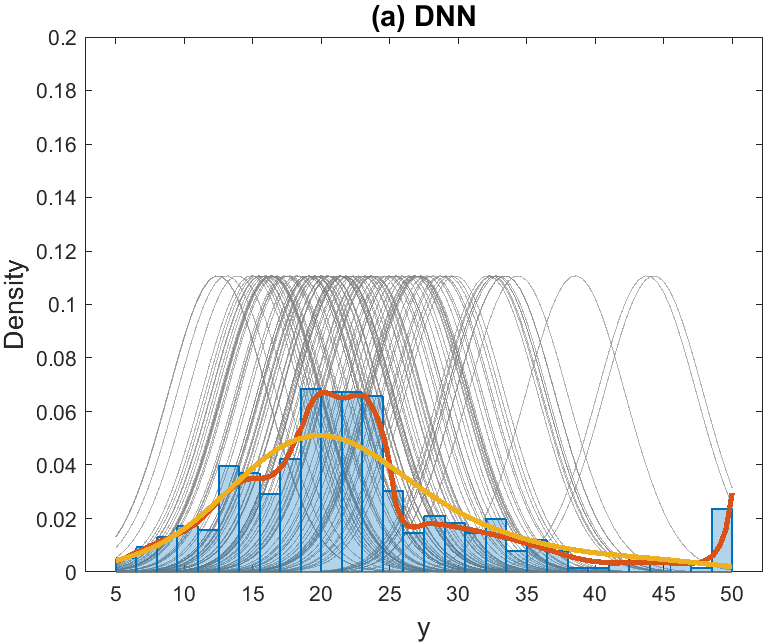}\hspace{0.09cm}
\centering\includegraphics[width=0.327\textwidth,angle=0]{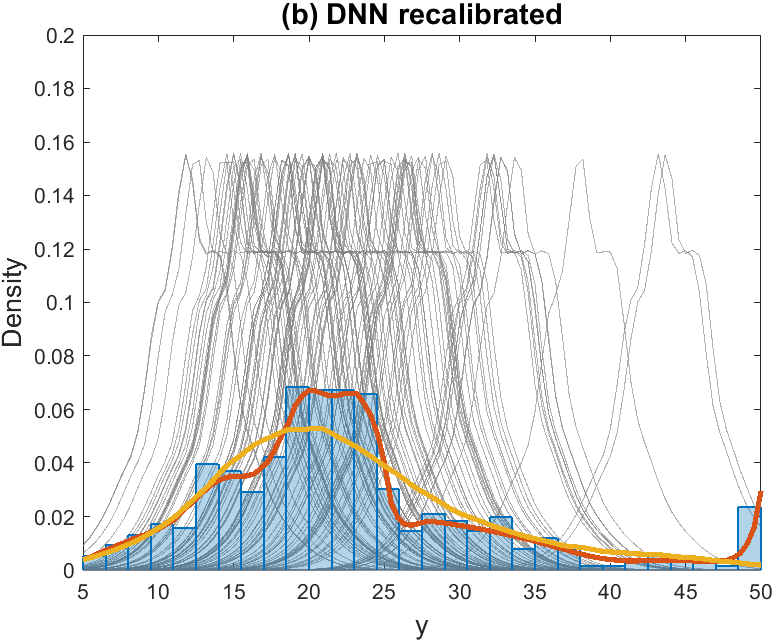}
\centering\includegraphics[width=0.327\textwidth,angle=0]{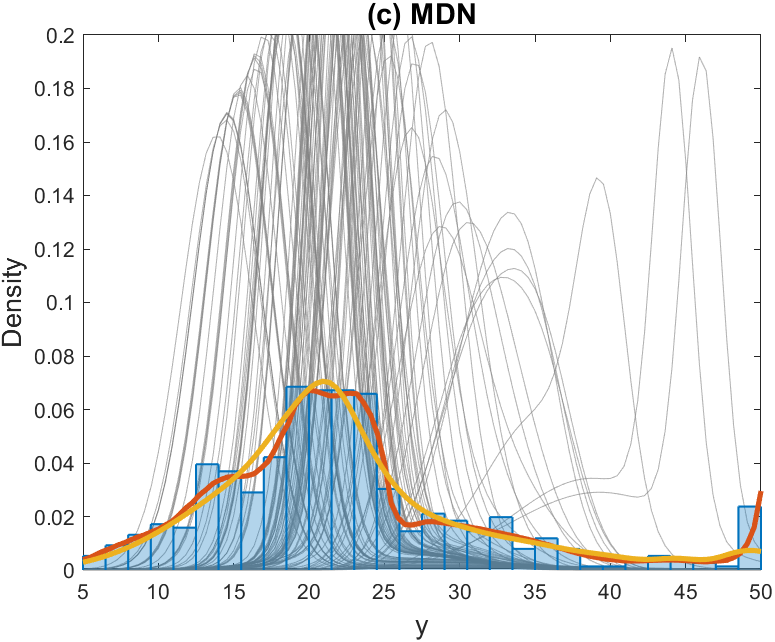}\\\vspace{0.1cm}
\centering\includegraphics[width=0.327\textwidth,angle=0]{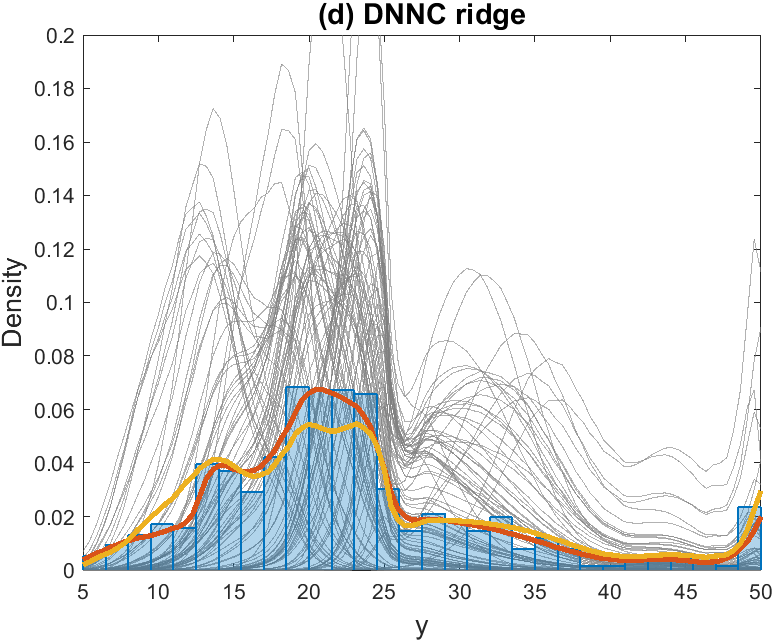}
\centering\includegraphics[width=0.327\textwidth,angle=0]{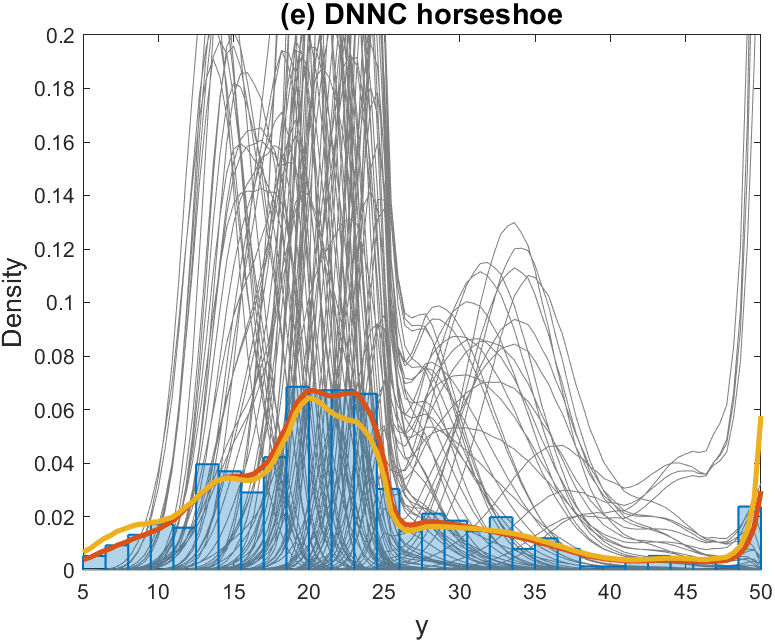}
\end{center}
Results are given for (a)~DNN, (b)~DNN-recalibrated, (c)~MDN, (d)~DNNC-ridge, and
(e)~DNNC-horseshoe. 
Each panel depicts predictive densities $\hat p_i(y|\xvec_i)$ (grey lines) 
for 100 of the $n=506$ observations, corresponding
to first 100 observations of the ordered response values. Also shown are a histogram (blue)
and KDE (red line)
of the response, 
and the average predictive density $\hat p_{\mbox{\tiny{marg}},n}(y)$ (yellow).  Accurate marginal
calibration is indicated by the red and yellow lines being very close.
\label{fig:Boston:densities}
\end{figure}

\begin{figure}[ht]
\caption{Predictive densities for the Cholesterol data.}
\begin{center}
\centering\includegraphics[width=0.325\textwidth,angle=0]{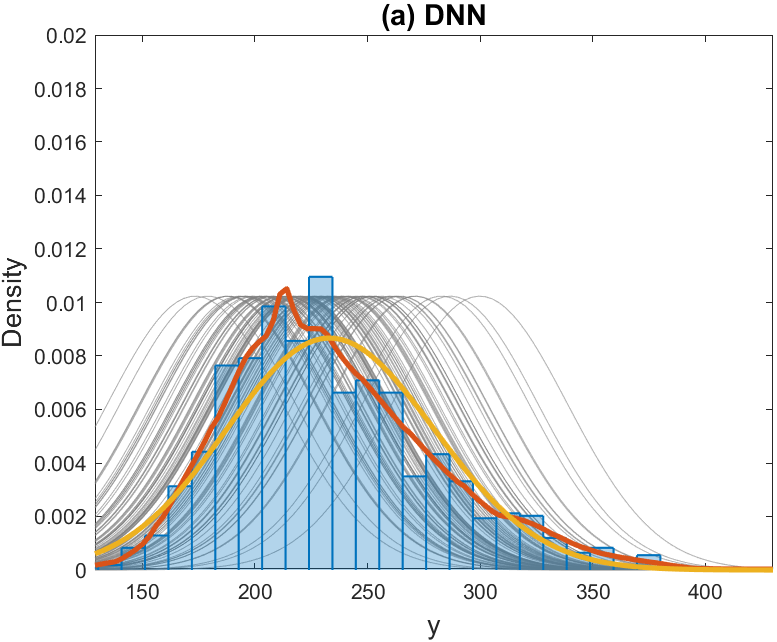}\hspace{0.02cm}
\centering\includegraphics[width=0.33\textwidth,angle=0]{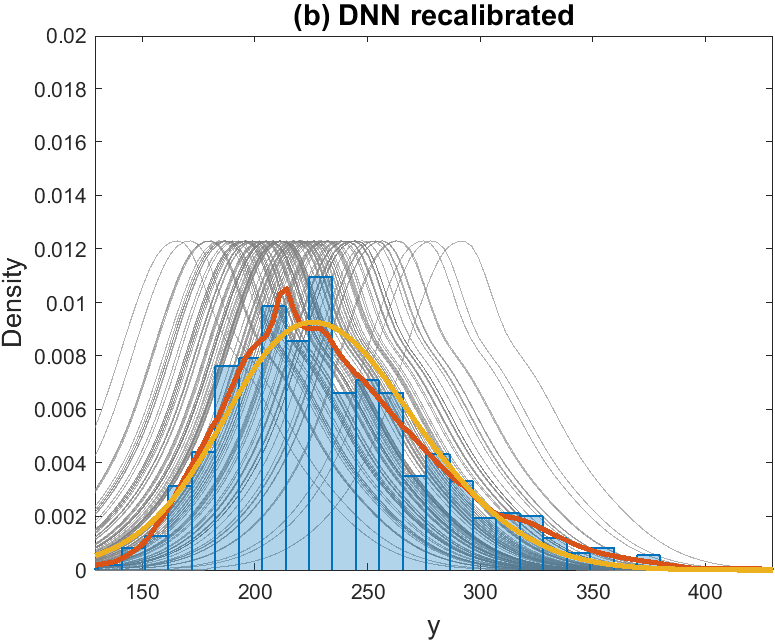}
\centering\includegraphics[width=0.327\textwidth,angle=0]{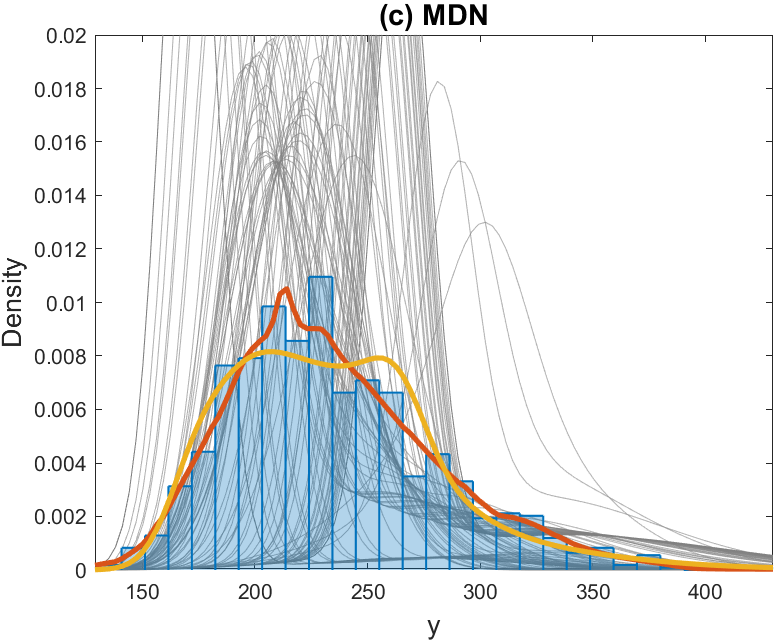}\\\vspace{0.1cm}
\centering\includegraphics[width=0.327\textwidth,angle=0]{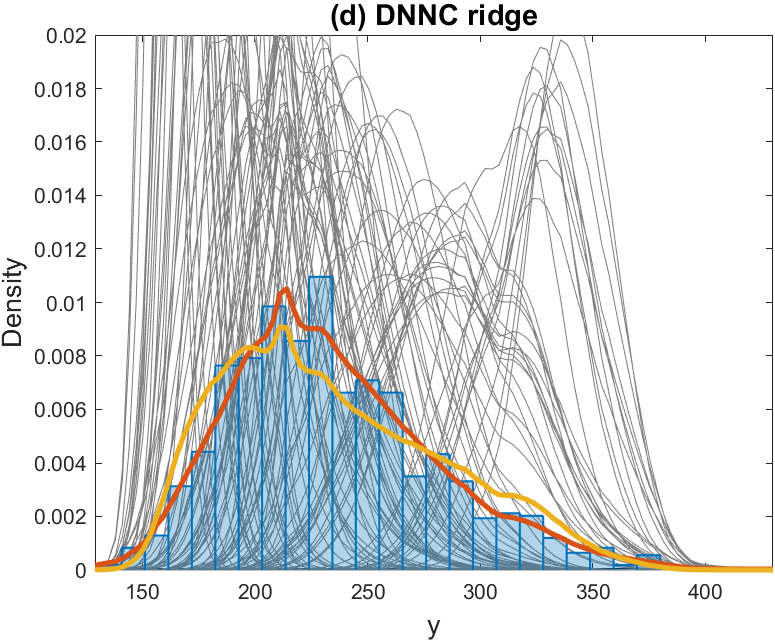}
\centering\includegraphics[width=0.327\textwidth,angle=0]{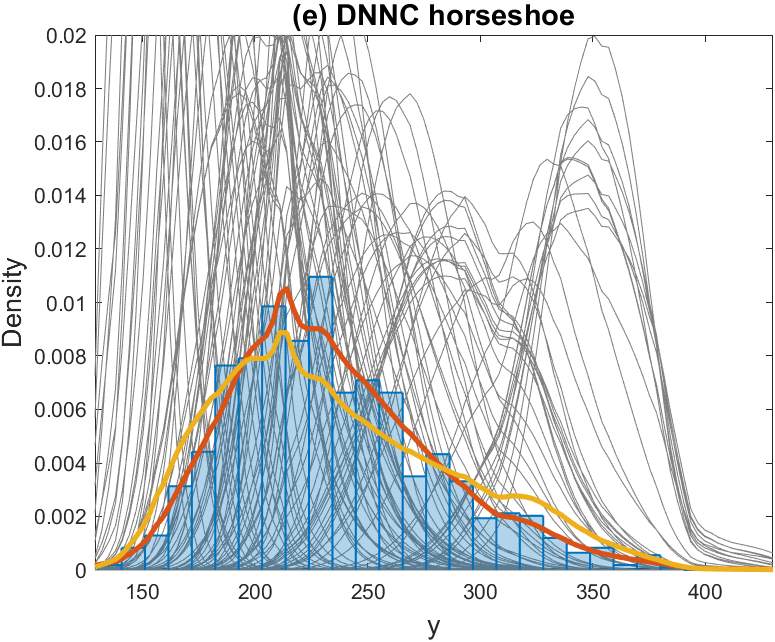}
\end{center}
Results are given for (a)~DNN, (b)~DNN-recalibrated, (c)~MDN, (d)~DNNC-ridge, and
(e)~DNNC-horseshoe. 
Each panel depicts predictive densities $\hat p_i(y|\xvec_i)$ (grey lines) 
for 100 of the $n=1,044$ observations, corresponding
to the first 100 observations of the ordered response values. Also shown are a histogram (blue)
and KDE (red line)
of the response, 
and the average predictive density $\hat p_{\mbox{\tiny{marg}},n}(y)$ (yellow). Accurate marginal
calibration is indicated by the red and yellow lines being very close.
\label{fig:chol:densities}
\end{figure}

\paragraph{Empirical results}
To first illustrate the difference in methods, Figures~\ref{fig:Boston:densities} and~\ref{fig:chol:densities} plot the predictive densities $\hat p_i(\cdot|\xvec_i)$,
evaluated at 
observations in the Boston housing and cholesterol datasets, respectively. 
For visual clarity, we only plot densities for observations that correspond
to the first 100 ordered response values in the data sets, resulting
in  100 predictive densities. Each panel corresponds to a different method,
and also includes a histogram of the response values and an adaptive kernel density
estimate (KDE). 
In panel~(a) the DNN produces homoscedastic Gaussian densities, 
whereas in panel~(b) the DNN-recalibrated method produces densities that 
are non-Gaussian, but still homoscedastic.
In contrast, the DNNC provides non-parametric predictions in panels~(d) and~(e) which are 
highly heteroscedastic.
As discussed in depth in~\cite{smith+k19}, a key 
strength of the regression copula modelling approach is that the entire 
predictive distribution---including
higher order moments---can vary with feature values. 
The MDN
method in panel~(c) produces predictive densities that are similar to the DNNC methods. 

\begin{figure}[ht]
\caption{Calibration plots for the Boston housing data.}
\begin{center}
\includegraphics[width=0.487\textwidth,angle=0]{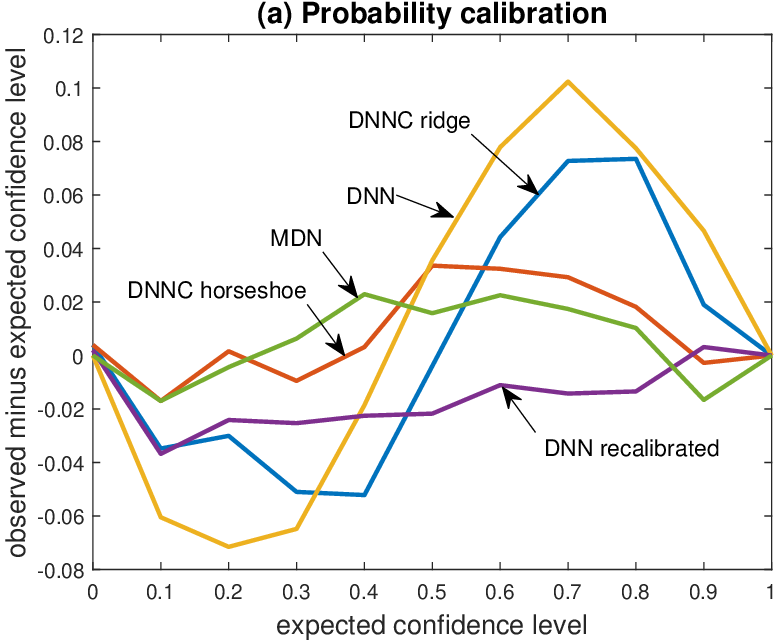}\hspace{0.19cm}
\includegraphics[width=0.49\textwidth,angle=0]{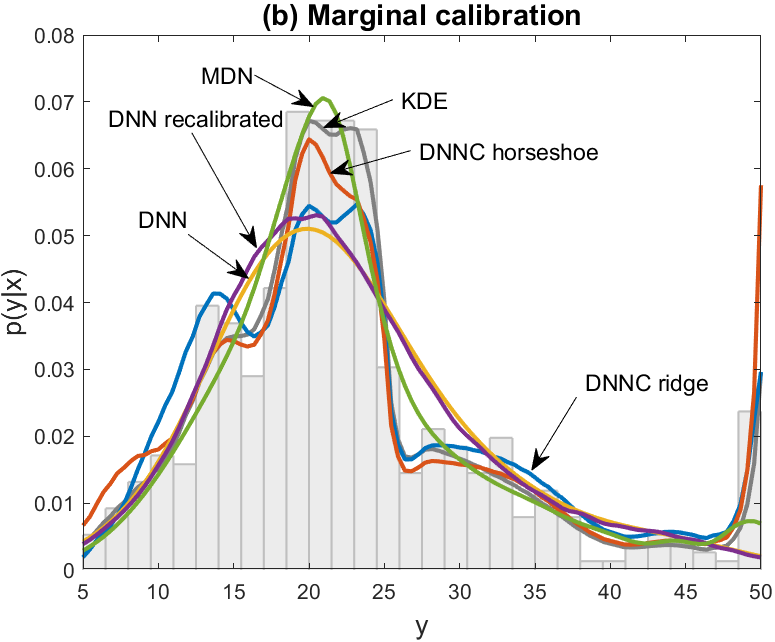}
\end{center}
Panel~(a) gives the probability calibration plot (with the y-axis showing difference between the observed and expected confidence levels to enhance visibility), and panel~(b) the
marginal calibration plot. Results are provided for the DNN (yellow line), DNN-recalibrated (violet line), MDN (green line)
DNNC-ridge (blue line), 
DNNC-horseshoe (red line). Also shown in~(b) is the 
kernel density estimate (grey line) and a histogram (grey) of the response.
\label{fig:Boston:cal}
\end{figure}

\begin{figure}[ht]
\caption{Calibration plots for the Cholesterol data.}
\begin{center}
\centering\includegraphics[width=0.49\textwidth,angle=0]{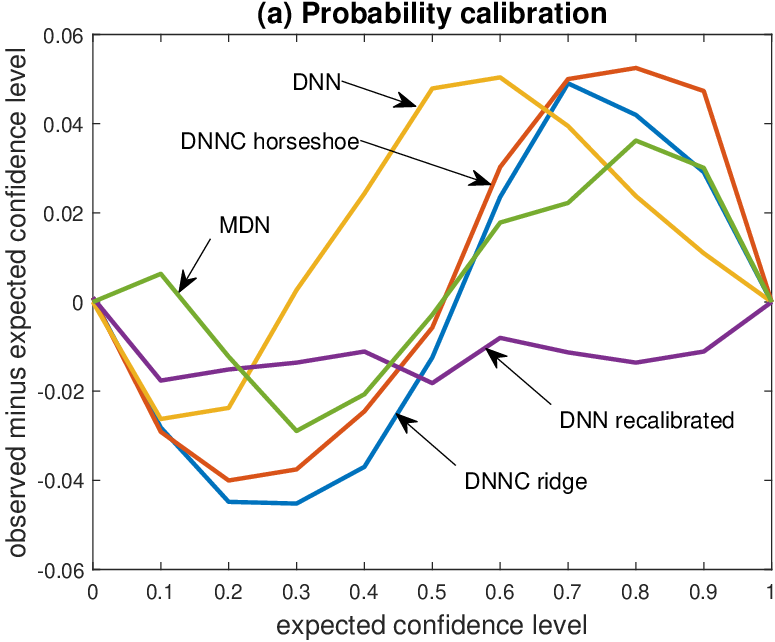}\hspace{0.05cm}
\centering\includegraphics[width=0.49\textwidth,angle=0]{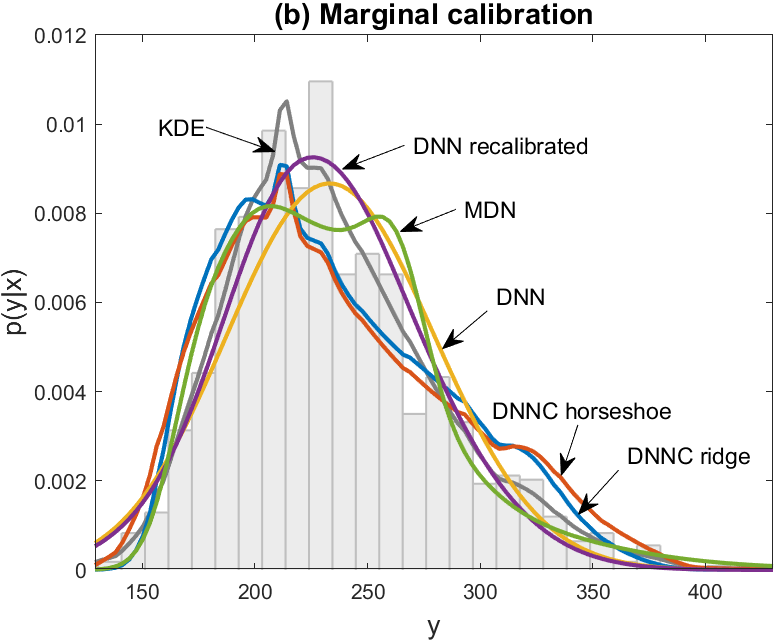}
\end{center}
Panel~(a) gives the probability calibration plot (with the y-axis showing difference between the observed and expected confidence levels to enhance visibility), and panel~(b) the
marginal calibration plot. Results are provided for the DNN (yellow line), DNN-recalibrated (violet line), MDN (green line)
DNNC-ridge (blue line), 
DNNC-horseshoe (red line). Also shown in~(b) is the 
kernel density estimate (grey line) and a histogram (grey) of the response.
	\label{fig:chol:cal}
\end{figure}

Figures~\ref{fig:Boston:cal} and~\ref{fig:chol:cal} show both 
calibration plots for the different methods and two datasets. We make four
observations. First, the DNN is neither marginally nor probability calibrated. 
Second, the DNN-recalibrated is probability calibrated well by construction,
but this does not lead to marginal calibration, which can be very poor, such as
in Figure~\ref{fig:Boston:cal}(b). 
Third, in contrast, the DNNC---whether using either the horseshoe
or ridge prior for regularization---exhibits accurate marginal calibration, along with near 
probability calibration. Fourth, the MDN method exhibits good probability calibration and
better marginal calibration than DNN and DNN-recalibrated, although not as good as the DNNC methods.


Finally, Table~\ref{tab:logsc} reports the in- and out-of-sample mean logarithmic scores
for the two datasets, and
we make three observations. First, the DNN performs poorly, which is because
the response distributions are non-Gaussian in both examples.  
Second, the shrinkage prior considered for $\betavec$ matters, with 
the predictions using the horseshoe prior superior to those using the ridge prior in both examples. 
Last, the DNNC is clearly more accurate 
than the benchmarks DNN and DNN-recalibrated, and slightly more than MDN. MDNs have
a universal approximation property which may be important in some problems.

\setlength{\abovedisplayskip}{0.1cm}
\setlength{\belowdisplayskip}{0.1cm}

\begin{table}
\renewcommand\arraystretch{1.2}
\begin{center}
\caption{Mean logarithmic scores for Boston and Cholesterol datasets.}
\label{tab:logsc}
\begin{tabular}{cccccc}
  \hline\hline
\multirow{ 2}{*}{Data} & DNN & DNN- & MDN  & DNNC- &  DNNC- \\
&  & recal. & & ridge & horseshoe  \\ \hline
&\multicolumn{5}{l}{{\em In-sample log-scores}}\\ \cline{2-6}
	Boston 	  & {-2.69}\footnotesize{(0.060)} &  -2.62\footnotesize{(0.056)}  &  {-2.39}\footnotesize{(0.049)} &  -2.39\footnotesize{(0.041)} &  \textbf{-2.17}\footnotesize{(0.041)}\\
	Chol. & -5.08\footnotesize{(0.027)} &  \textbf{-4.22}\footnotesize{(0.030)} & -4.65\footnotesize{(0.034)} & -4.35\footnotesize{(0.026)} & {-4.31}\footnotesize{(0.026)}\\
 &\multicolumn{5}{l}{{\em Predictive log-scores}}\\ \cline{2-6}
  Boston  &  -2.84\footnotesize{(0.084)}  &  -2.61\footnotesize{(0.047)} &   -2.61\footnotesize{(0.065)} & -2.71\footnotesize{(0.056)}  & \textbf{-2.56}\footnotesize{(0.056)}\\  
  Chol. & -5.14\footnotesize{(0.030)}  & -5.09\footnotesize{(0.029)}  & -5.19\footnotesize{(0.047)} & \textbf{-5.05}\footnotesize{(0.030)} & { -5.07}\footnotesize{(0.032)}\\
  \hline\hline
\end{tabular}
\end{center}
Scores for each of the five methods are given in the columns, and for both in-sample (top half) and out-of-sample (bottom half) predictions. 
The latter are constructed using ten-fold cross-validation, as discussed in the 
text. Higher values correspond to more accurate predictive densities, with the highest
values for each case in bold. Standard errors for the means are 
also given in parentheses. 
\end{table}

\section{Application to Likelihood-Free Inference}\label{sec:LFI}
In this section we discuss likelihood-free inference.  We will discuss here only regression approaches
to likelihood-free inference.  For a recent comprehensive overview of alternative methods see 
\citet{sisson+fb18}.
We show how to use the distributional deep regression copula model
in Section~\ref{sec:DNN} to perform
likelihood-free inference, and highlight the advantage of marginal calibration---which is an intrinsic aspect of the copula model---in this context.
To illustrate, a regression copula is constructed from a
convolutional network, and the resulting copula model is used to construct
likelihood-free inference in two empirical applications.  Convolutional networks are used in situations where the features
take the form of a time series or an image, and the layers of the network can be thought of as extracting local characteristics of the input
and then combining these into increasingly abstract representations.  See~\cite{polson+s17} and~\cite{fan2019} for further background.  

\subsection{Likelihood-free inference}

\subsubsection{Introduction}
Let $\rhovec$ denote the parameters
in a parametric statistical model for data observed from
a sampling distribution with density $p(\dvec|\rhovec)$.
Consider Bayesian inference with prior density $p(\rhovec)$,
and 
observed data denoted as $\dvec_{\text{obs}}$, so that the posterior  
density is $p(\rhovec|\dvec_{\text{obs}})$.\footnote{We denote the data vector as $\dvec$, rather than $\yvec$, to 
	avoid confusion with the response values from the distributional deep regression
	copula model outlined in Section~\ref{sec:DNN}.}
Suppose we can simulate data $(\rhovec_i,\dvec_i)$, $i=1,\dots, n$ as $\rhovec_i\sim p(\rhovec)$, $\dvec_i\sim p(\dvec|\rhovec_i)$. We refer to the density
$p(\rhovec,\dvec)=p(\rhovec)p(\dvec|\rhovec)$ as the `joint model' for the data and parameters. By definition, the posterior density is the conditional density of $\rhovec$ given $\dvec=\dvec_{\text{obs}}$ obtained from this joint
density.
In cases where the likelihood function $p(\dvec|\rhovec)$ is intractable (but the model can
still be simulated from),
we can use a regression
fitted to the simulated data to approximate the conditional density of $\rhovec$ given $\dvec$ in
the joint model. Fitting a regression model
 to data $(\rhovec_i,\dvec_i)$, $i=1,\ldots,n$, where $\rhovec_i$ is the response and $\dvec_i$ is the feature vector,
will give a predictive
density $\tilde{p}(\rhovec|\dvec)$ for any $\dvec$. The predictive density $\tilde{p}(\rhovec|\dvec_{\text{obs}})$ is then an approximation to the posterior 
density based on the regression model. Thus, selecting a regression method
that produces an accurate density estimate $\tilde{p}(\rhovec|\dvec)$ is key 
to conducting accurate likelihood-free inference in this approach.  

Later we consider only modelling scalar functions of $\rhovec$ using separate regressions, rather than a multivariate regression model. Posterior distributions for one-dimensional
functions of the parameter are enough for scientific inferences in many cases, 
although the joint posterior is needed for some purposes.  Predictive inference requires the full joint posterior, and
even for scientific inferences a joint posterior on several parameters may sometimes be required. For example, \cite{wood10} 
considers the assessment of different dynamic regimes for the blowfly data discussed in Section~\ref{sec:CNNegs}, which are dependent
on several of the model parameters. Multivariate extensions of our methods are left to future work.

\subsubsection{Advantage of marginal calibration}
For simplicity, we denote scalar functions of $\rhovec$ as $\rho$. Then the marginal distribution 
for the regression training data $\rho_i\sim p(\rho)$, $i=1,\dots, n$ is the prior $p(\rho)$.  
In this case, marginal calibration
as defined in Section~\ref{sec:uncertDNN} occurs when 
the marginal distribution for $\rho$ (i.e. $p(\rho)$ here) matches
the average posterior predictive density $\frac{1}{n} \sum_{i=1}^n \tilde{p}(\rho|\dvec_i)$ from the regression model. Moreover,
this average is a sample-based estimate of $\int \tilde{p}(\rho|\dvec) p(\dvec)\mbox{d}\dvec$, 
because the values $\dvec_i$ are simulated from density
$p(\dvec)=\int p(\rho)p(\dvec|\rho)\,d\rho$.
Therefore, 
if the marginal calibration property holds for the regression method then we have
\begin{align}
\frac{1}{n} \sum_{i=1}^n \tilde{p}(\rho|\dvec_i) & \approx \int \tilde{p}(\rho|\dvec) p(\dvec)\mbox{d}\dvec=p(\rho).  \label{calibration1}
\end{align}
To highlight why it is advantageous for such a property to hold, write the joint Bayesian model as $p(\rho,\dvec)=p(\dvec)p(\rho|\dvec)$ and then notice that
\begin{align}
p(\rho) & = \int p(\rho|\dvec)p(\dvec) \mbox{d}\dvec\,, \label{calibration2}
\end{align}
which shows that marginal calibration always holds for the true posterior density. 
Thus it is desirable for the regression-based approximations to the 
posterior to respect this kind of calibration also.  However, while marginal calibration is a necessary quality for a 
regression approximation of the posterior to be good, it is not sufficient. 
For example, if we always estimate the posterior density by the prior regardless of the data, then this 
approximation method is marginally calibrated but does not give good posterior approximations. 
Nevertheless, we show empirically in our later examples that our copula method achieves both better marginal calibration and uncertainty quantification than benchmark methods.  

\subsubsection{Previous flexible regression models for likelihood-free inference}
The use of flexible regression models for conditional density estimation in likelihood-free inference is not new.  For example, \citet{fan+ns13} consider flexible regression
models for approximating the summary statistic distribution, and hence the likelihood, based on a copula of a mixture and flexible mixture of experts
regression estimates of summary statistic marginal distributions.  \citet{raynal+mprre18} consider applying the quantile regression
forests method of \citet{meinshausen06} to likelihood-free inference.  \citet{izbicki+l17} describe methods for converting 
high-dimensional regression methods into flexible conditional density estimators using orthogonal series estimators, 
and \citet{izbicki+lp19} consider initial estimates obtained from an ABC sampler, and then applying non-parametric conditional density
estimators which make use of a surrogate loss function to estimate the conditional density locally. \citet{papamakarios+m16} consider a neural network approach using
a sequential decomposition of the posterior into conditional distributions and mixture density network models for the conditionals.  They 
also consider the use of a sequential design strategy to concentrate more on the high posterior probability region of the parameter space.  
Further improvements on this methodology are given in \citet{lueckmann+gbonm17}.  
\citet{papamakarios+sm19} consider autoregressive flows to learn approximations to the likelihood, 
avoiding the difficulties of removing the bias arising from the use of a proposal in the sequential design
step in \citet{papamakarios+m16} and \citet{lueckmann+gbonm17}.  However, methods such as those of \citet{papamakarios+sm19} and 
\citet{fan+ns13} require use of a
conventional Bayesian computational algorithm for summarizing the posterior once the approximate likelihood has been obtained.  
Heteroscedastic neural network methods have also been used in post-processing adjustments of
conventional ABC samplers \citep{blum+f10}, where empirical residuals 
are used within the fitted regression model to give particle approximations to the posterior.  The method of \citet{blum+f10} 
builds on the earlier seminal paper of~\citet{beaumont+zb02}. 

Many of the regression methods discussed above require the use of summary statistics for their application. In contrast, our 
approach does not require summary statistics, and makes use of convolutional neural networks (CNNs) for specifying the marginal posterior
distributions, where the time series values are used directly as features of the CNNs in our copula-based distributional regression model. \citet{jiang+wzw17} is the first work,  of which we are aware, that uses deep learning methods
to automate the choice of summary statistics in likelihood-free inference, although they do not consider convolutional networks.
Convolutional networks have been used in likelihood-free inference for time series by \citet{dinev+g18}, where similar to
\citet{jiang+wzw17} the regression is being used as an automated
way of obtaining summary statistics, rather than for density estimation in itself. \citet{greenberg+nm19} suggest a way 
to use a proposal which focuses on a relevant part of the parameter space without the difficulties of the proposal corrections 
in \citet{papamakarios+m16} and \citet{lueckmann+gbonm17}, and also avoiding additional computations
after the conditional density estimation step.  
In their method, a parametrized family of approximations is considered, such as Gaussian or a mixture of Gaussians, 
and a mapping from the data to the parameters
in the approximation is learned using a certain loss function.  In the case of time series data, it may be possible to avoid the use of summary
statistics using this approach for a suitable neural network parametrization of the function mapping the data to the parameters of the approximation.  
The general principle of marginal calibration can be applied in conjunction with some of the other flexible regression methods
for specification of the all aspects of the posterior distribution 
described above, 
complementing the existing literature on 
flexible regression methods in likelihood-free inference.  

\subsection{Convolutional network examples}\label{sec:CNNegs}
In the context of likelihood-free inference for time series models, \citet{dinev+g18} 
considered using a CNN regression model to predict the components of the parameter
vector $\rhovec\in\dsR^p$ based on data $\dvec=(d_1,\ldots,d_T)^\top$, using a training set of simulations of pairs $(\rhovec_k,\dvec_k)$ which are generated
from the joint model. \citet{dinev+g18} consider multivariate outputs to predict all parameters jointly and use the 
predictions of the network as an automated summary statistic choice.  
Here we will use the DNN regression copula model in Section~\ref{sec:DNN}
as a distributional 
regression method to directly estimate the marginal posterior distributions 
of the elements of $\rhovec$.

We show the potential of our proposed approach (labelled DNNC) in two complex ecological time
series examples. In both examples we simulate 10000 data sets  under the prior, and use 8000 data sets for training and 2000 for testing, denoted as $\lbrace \dvec_i\rbrace$ and $\lbrace \dvec_j^{\ast}\rbrace$, respectively. Each simulated series has the same length $T$ as the observed data. In the simulation, the accuracy of parameter predictions 
can be measured directly for different likelihood-free methods. 
For the real data, we can split the data and 
measure the accuracy of the parameter point estimates using a composite scoring rule for the different methods.

Although the whole time series is used as the feature vector in our regression models, our methodology scales well
with $T$. CNNs can be trained easily even for long series using 
standard deep learning libraries. Additionally, because of the way CNNs extract local information from the
series (by applying filters with fixed weights, and then combining this information at different scales), 
the number of weights to be learned does not grow rapidly with $T$ for suitable network architectures.  

\paragraph{Nicholson's blowfly model}
As the first example we consider the data reported by~\citet{Nic1954}
from laboratory experiment E2 to elucidate the population
dynamics of sheep blowfly (\emph{Lucilia cuprina}). 
\cite{wood10} modelled the observed dynamics of the population
at time $t$ as $n_t=r_t+s_t$, where
%
\[
r_t\sim\mbox{Poisson}\left(Pn_{t-\tau}\exp(-n_{t-\tau}/n_0)e_t\right)\,,
\]
is the delayed recruitment process with parameters $P$, $n_0$, $\tau$, $\delta$, and
\[
s_t\sim\mbox{Binomial}(\exp(-\delta\epsilon_t),n_{t-1}),
\]
is the adult survival process. Here, $e_t$, $\epsilon_t$ are independent gamma disturbances with unit mean and variances $\sigma_p^2$ and $\sigma_d^2$, respectively. Consequently, $p=6$, $\rhovec=(\delta,P,n_0,\sigma_p^2,\tau,\sigma_d^2)$ and the length of the series is $T=275$. We use the prior for $p(\rhovec)$ specified in Table~13 of~\citet{FasWoo2016} and given in
the Web Appendix.

\paragraph*{A chaotic prey-predator model for modelling voles abundance} The second model we consider is used by \citet{SisFanBea2018} to describe the dynamics of Fennoscandian voles (\emph{Microtus and Clethrionomys}). There has been an observed shift in voles abundance dynamics from low-amplitude oscillations in central Europe and southern Fennoscandia
to high-amplitude fluctuations in the north.  One possible reason is the
absence of generalist predators in the north, where voles are hunted primarily by weasels
(\emph{Mustela nivalis}). In the notation of \citet{SisFanBea2018}, the predator-prey dynamics are given by the following system of differential equations~\citep{TurEll2000}
\begin{equation*}\begin{aligned}
\frac{dN}{dt} & = r(1-e\sin(2\pi t))N-\frac{r}{K}N^2-\frac{GN^2}{N^2+H^2}-\frac{CNP}{N+D}+\frac{N}{K}\frac{dw}{dt}\\
\frac{dP}{dt} & = s(1-e\sin(2\pi t))P-sQ\frac{P^2}{N},
\end{aligned}\end{equation*}
where  $dw(t_2)-dw(t_1)\sim\ND(0,\sigma^2(t_2-t_1))$, $t_2>t_1$, is a Brownian motion process with constant volatility $\sigma$; and $N$ and $P$ represent voles and weasels abundances, respectively.  \citet{TurEll2000} considered a mechanistic version of this model without the driving Brownian motion, and investigate the effects of environmental noise through perturbing the model parameters.  The model is formulated in continuous time, with the parameters $r$ and $s$ representing intrinsic population growth rates of voles and weasels respectively, while $K$ is the carrying capacity of $r$. Averaging of these parameters is done over the seasonal component with amplitude $e$ and period equal to one year.  Peak growth is achieved in summer. Further, $G$ and $H$ as well as $C$ and $D$ are the parameters of type II and III functional response models of generalist predation and predation by weasels, respectively; see~\citet{SisFanBea2018} for further details. These authors assume
the number of trapped voles to be Poisson distributed, $d_t\sim\mbox{Poisson}(\Phi N_t)$, at times 
$t\in\lbrace 1,\ldots, T\rbrace$ when trapping took place.  Finally, the model is not fitted directly to data but rescaled to a dimensionless form, where
\[
n=\frac{N}{K},\;p=\frac{QP}{K},\;\delta=\frac{D}{K},\;a=\frac{C}{Q},\;g=\frac{G}{K},\;h=\frac{H}{K},\;\phi=\Phi K,
\]
and
\begin{equation*}\begin{aligned}
\frac{dn}{dt} & = r(1-e\sin(2\pi t))n-rn^2-\frac{gn^2}{n^2+h^2}-\frac{anp}{n+\delta}+n\frac{dw}{dt}\\
\frac{dp}{dt} & = s(1-e\sin(2\pi t))p-s\frac{p^2}{n}\\
d_t & \sim\mbox{Poisson}(\phi n_t).
\end{aligned}\end{equation*}
In this dimensionless form of the model we continue to write $\sigma$ for the Brownian motion scale parameter, although it is not the same parameter in the two models.  We used the same strategy as~\citet{SisFanBea2018} to arrive at $T=90$ data points collected during the spring (mid-June) and autumn (September) of each year between 1952 and 1997. The priors on the parameter $\rhovec=(r,e,g,h,a,\delta,s,\sigma,\phi)^\top$  are specified in 
the Web Appendix.

\paragraph*{CNN architecture}
To set up the basis functions for our DNNC in Step~2 of Algorithm~1, 
we fitted separate CNNs for each model parameter (six parameters for the blowfly model, nine for the voles abundance model) with pseudo-responses $\zvec$ as outputs and the simulated series as features. 
Beginning with the generic architecture suggested in \citet{dinev+g18} and
experimenting also with other specifications, we ended up using two convolutional layers with 31/7 filters, and kernel sizes of 31/10 for the blowfly and voles examples respectively.  
In the later cross-validation comparisons where training based on the first 80\% of the series is considered, the kernel sizes were 31/8 for the blowfly and voles data, respectively.
We used ReLU activation functions for the convolutional layers with L2 regularization with parameter 0.001, followed by two dense layers of sizes 100/1 and ReLU/linear activation functions for the first and second dense layers respectively. It is important to not normalize the inputs as this may destroy the time series structure. Instead we use batch normalization after each layer. The number of epochs was determined by cross-validation and a batch-size of 256. As before we employed the optimizer adam with default settings.
As proposed by \citet{dinev+g18} we apply a max-pooling layer after the first convolutional layer and a flattening layer after the second one. Extracting the resulting basis functions was done as in the previous subsection using the \texttt{keras} package.

\paragraph*{Benchmark models}
To compare our DNNC method with, we consider the following benchmark methods, with methods~(iii) to~(vi) being leading approaches in likelihood-free inference: 
\begin{itemize}
\item[(i)] \underline{DNN}:uses the same CNN architecture as employed to construct the copula, but applied directly to the
original response values, and with Gaussian errors in the output layer.
\item[(ii)] \underline{DNNCss}: uses summary statistics as features in the regression and the copula methodology
of Section 4 with a two hidden layer FNN.
\item[(iii)] \underline{ABC}: the method of  \citep{blum+f10} with a
DNN regression adjustment, as implemented in the R package \texttt{abc} by \citet{csillery+fb12}. 
\item[(iv)] \underline{ABCrf}: ABC with random forests, as implemented in the R package \texttt{abcrf} by \citet{marin+rpre17}. 
\item[(v)] \underline{BSL}: the Bayesian synthetic likelihood approach implemented in the R package \texttt{BSL} by \citet{an+sd19}). 
\item[(vi)] \underline{semiBSL}: a semi-parametric version of BSL which estimates a Gaussian copula model with non-parametric marginal distributions,
as implemented in the R package \texttt{BSL}. 
\end{itemize}
The methods DNNCss, BSL, semiBSL, ABC and ABCrf require summary statistics, and we used the same choices as~\citet{FasWoo2016} and~\citet{SisFanBea2018} (23 statistics for the blowfly data and 16 for the voles data) and implemented in the R packages \texttt{synlik}~\citep{fasiolo+w14} and \texttt{volesModel} (available on Github). Most of these methods require careful
implementation, and we give extensive details on these
in Appendix~\ref{appA}, and additional comments on the computational
demands of each approach in the Web Appendix.
Due to the high computational cost we exclude BSL and semiBSL from the simulation but use them for the two real data analyses.
\vspace{-10pt}

\begin{table}[tbh]
	\begin{center}
		\caption{Simulation results for the blowfly test data sets.}\label{tab:sim:cnn}
		{\small
		\begin{tabular}{c|ccccc}
			\hline\hline
			& ABC & ABCrf & DNN & DNNC & DNNCss\\\hline
			$\delta$ & 1.35 \footnotesize{(0.046/0.50)} & 1.30 \footnotesize{(0.047/0.48)} & 1.21 \footnotesize{(0.043/0.62)} & 1.02 \footnotesize{(0.037/0.75)} & 1.32 \footnotesize{(0.043/0.66)}\\
			$P$ & 0.48 \footnotesize{(0.014/0.82)} & 0.50 \footnotesize{(0.015/0.77)}  & 0.41 \footnotesize{(0.010/0.87)} & 0.33 \footnotesize{(0.013/0.89)} & 0.46 \footnotesize{(0.014/0.88)}\\ 
			$n_0$  &  1.38 \footnotesize{(0.054/0.54)} &  1.41 \footnotesize{(0.056/0.44)} & 1.33 \footnotesize{(0.035/0.77)} & 0.80 \footnotesize{(0.050/0.95)} & 1.35 \footnotesize{(0.052/0.77)}\\
			$\sigma_p^2$ &  1.44 \footnotesize{(0.063/0.74)} &  1.54 \footnotesize{(0.064/0.67)} & 1.05 \footnotesize{(0.050/0.90)} & 0.96 \footnotesize{(0.051/0.94)} & 1.41 \footnotesize{(0.059/0.85)}\\
			$\tau$ & 0.29 \footnotesize{(0.018/0.70)} & 0.31 \footnotesize{(0.021/0.66)}  & 0.20 \footnotesize{(0.016/0.92)} & 0.21 \footnotesize{(0.016/0.96)} & 0.27 \footnotesize{(0.017/0.88)}\\
			$\sigma_d^2$ & 1.11 \footnotesize{(0.049/0.87)} & 1.14 \footnotesize{(0.050/0.86)} & 0.92 \footnotesize{(0.043/0.95)}  & 0.88 \footnotesize{(0.041/0.95)} & 1.07 \footnotesize{(0.046/0.90)}\\
			\hline\hline
		\end{tabular}
	}
	\end{center}
	Each cell gives the mean squared error for the logarithm of the parameters, along with its standard error and coverage
	of the 95\% credible interval in parentheses.
	Each row corresponds to a different parameter, while the columns give results for the ABC, ABCrf, DNN, DNNC and DNNCss methods. 
\end{table}

\begin{table}[tbh]
	\begin{center}
		\caption{Simulation results for the voles test data sets.}\label{tab:sim:cnn:voles}
		{\small
		\begin{tabular}{c|ccccc}
			\hline\hline
			& ABC & ABCrf & DNN & DNNC &DNNCss\\\hline
			$r$ &  0.04 \footnotesize{(0.002/0.94)}  & 0.02 \footnotesize{(0.001/0.99)} & 0.03 \footnotesize{(0.001/0.92}) & 0.03 \footnotesize{(0.001/0.94)} & 0.03 \footnotesize{(0.001/0.97)}\\
			$e$ & 0.81 \footnotesize{(0.046/0.95)} & 0.16 \footnotesize{(0.010/0.99)} & 0.24 \footnotesize{(0.014/0.93)} & 0.19 \footnotesize{(0.017/0.94)} & 0.34 \footnotesize{(0.024/0.97)}\\
			$g$ & 1.56 \footnotesize{(0.069/0.95)}  & 1.36 \footnotesize{(0.060/0.97)}  & 1.23 \footnotesize{(0.053/0.94)} & 1.17 \footnotesize{(0.057/0.95)} & 1.36 \footnotesize{(0.060/0.95)}\\
			$h$ &  0.27 \footnotesize{(0.009/0.95)} &  0.25\footnotesize{ (0.008/0.96)} & 0.23 \footnotesize{(0.008/0.94)}& 0.22 \footnotesize{(0.008/0.95)} & 0.23  \footnotesize{(0.001/0.95)}\\
			$a$ &  0.92 \footnotesize{(0.065/0.95)} &  0.39 \footnotesize{(0.025/0.99)}   & 0.39 \footnotesize{(0.031/0.93)} & 0.36 \footnotesize{(0.037/0.95)} & 0.47  \footnotesize{(0.042/0.97)}\\
			$\delta$ & 0.60 \footnotesize{(0.023/0.95)} & 0.32 \footnotesize{(0.013/0.98)}  & 0.33 \footnotesize{(0.014/0.92)} & 0.32 \footnotesize{(0.014/0.94)} & 0.36  \footnotesize{(0.015/0.96)}\\
			$s$ & 0.59 \footnotesize{(0.032/0.94)} & 0.19 \footnotesize{(0.014/0.98)}  & 0.19 \footnotesize{(0.018/0.94)} & 0.17 \footnotesize{(0.022/0.95)} & 0.18  \footnotesize{(0.027/0.96)}\\
			$\sigma$ & 0.72 \footnotesize{(0.030/0.94)} & 0.21 \footnotesize{(0.006/0.99)}   & 0.30 \footnotesize{(0.016/0.96)}  & 0.26 \footnotesize{(0.015/0.93)} & 0.23 \footnotesize{(0.013/0.96)}\\
			$\phi$ & 0.32 \footnotesize{(0.073/0.95)} & 0.36 \footnotesize{(0.016/0.99)} & 0.41 \footnotesize{(0.034/0.97)}  & 0.34 \footnotesize{(0.037/0.95)} & 0.35 \footnotesize{(0.034/0.96)}\\
			\hline\hline
		\end{tabular}
	}
	\end{center}
	Each cell gives the mean squared error for the logarithm of the parameters, along with its standard error and coverage
	of the 95\% credible interval in parentheses.
	Each row corresponds to a different parameter, while the columns give results for the ABC, ABCrf, DNN, DNNC and DNNCss methods.
\end{table}

\begin{sidewaysfigure}[p]
\caption{Predictive marginal posteriors for the simulated test data from the blowfly model.}\label{fig:marg:CNN:sim}
\begin{center}
\includegraphics[width=0.99\textwidth,angle=0]{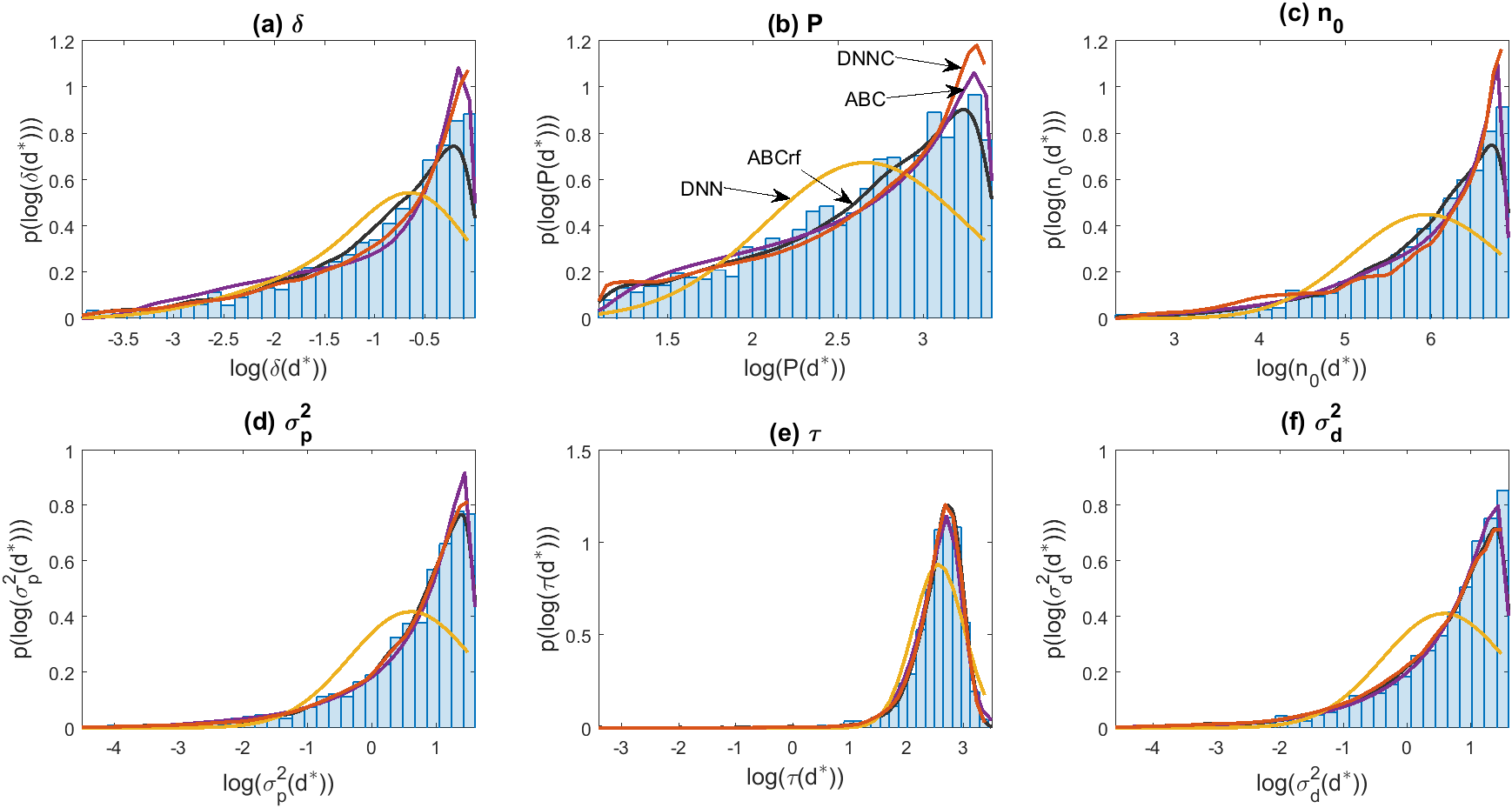}
\end{center}
Panels (a) to (f) show the average marginal predictive densities for the four methods  ABC (violet line), ABCrf (gray line), DNN (yellow line), DNNC (red line). Also shown in blue is a histogram of the simulated test data.
\end{sidewaysfigure}

\begin{sidewaysfigure}[p]
\caption{Predictive marginal posteriors for the simulated test data from the voles model.}\label{fig:marg:CNN:sim:voles}
\begin{center}
\includegraphics[width=0.99\textwidth,angle=0]{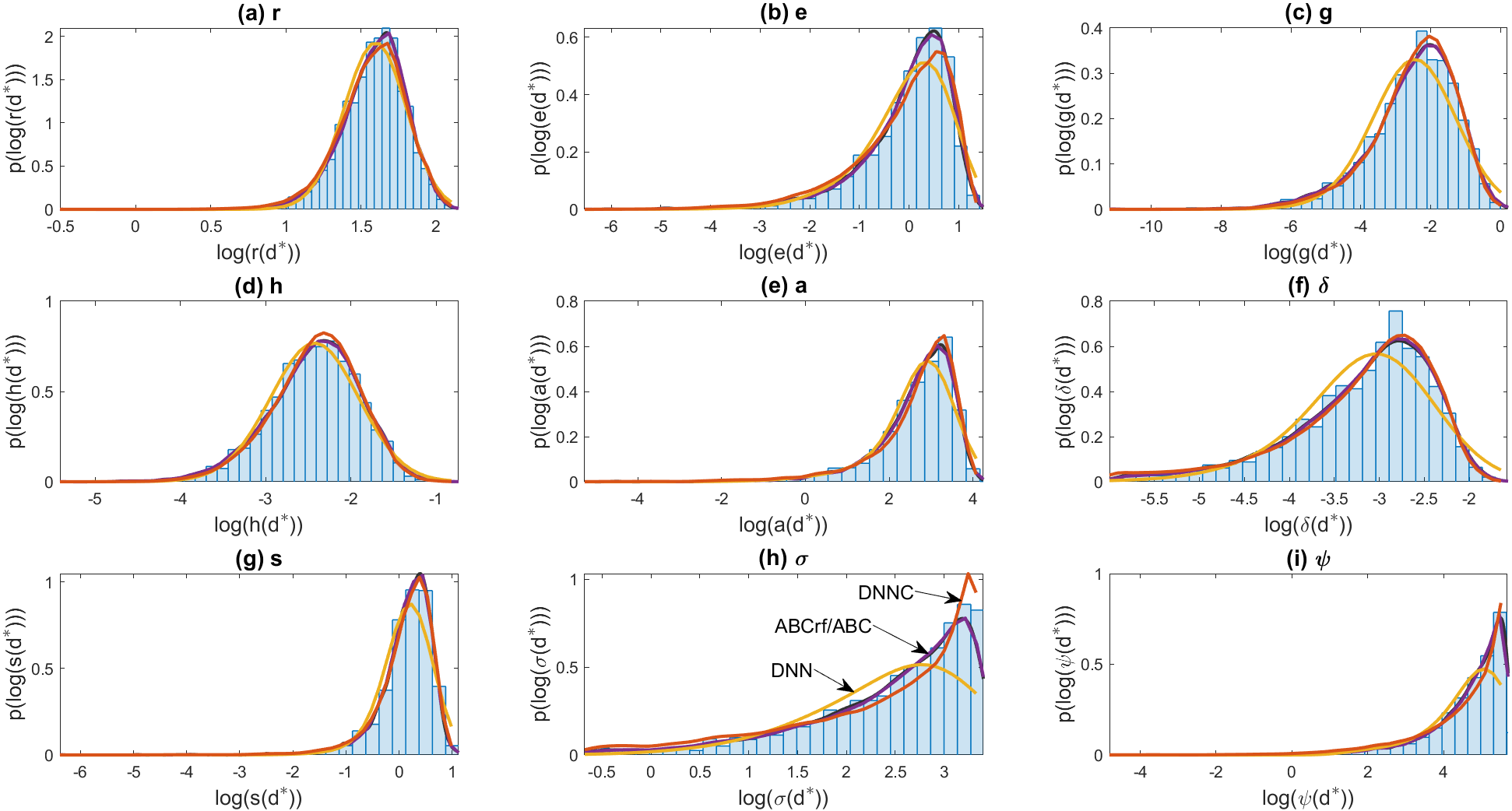}
\end{center}
Panels (a) to (f) show the average marginal predictive densities for the four methods  ABC (violet line), ABCrf (gray line), DNN (yellow line), DNNC (red line). Also shown in blue is a histogram of the simulated test data.
\end{sidewaysfigure}

\paragraph*{Measures of performance}
Tables~\ref{tab:sim:cnn} and \ref{tab:sim:cnn:voles} report mean squared errors, standard errors and coverage rates of 95\% credible intervals (in parentheses the latter two) for the logarithm of the parameters in the simulated test data sets for both applications. Figures~\ref{fig:marg:CNN:sim} and \ref{fig:marg:CNN:sim:voles} depict the average marginal posteriors of the log-parameters averaging over the 2,000 test replicates, to examine marginal calibration of different methods. 

\begin{sidewaysfigure}[p]
	\caption{Predictive marginal log-posteriors for the observed test data from the blowfly model.}
	\label{fig:marg:CNN:obs}
	\begin{center}
		\includegraphics[width=0.99\textwidth,angle=0]{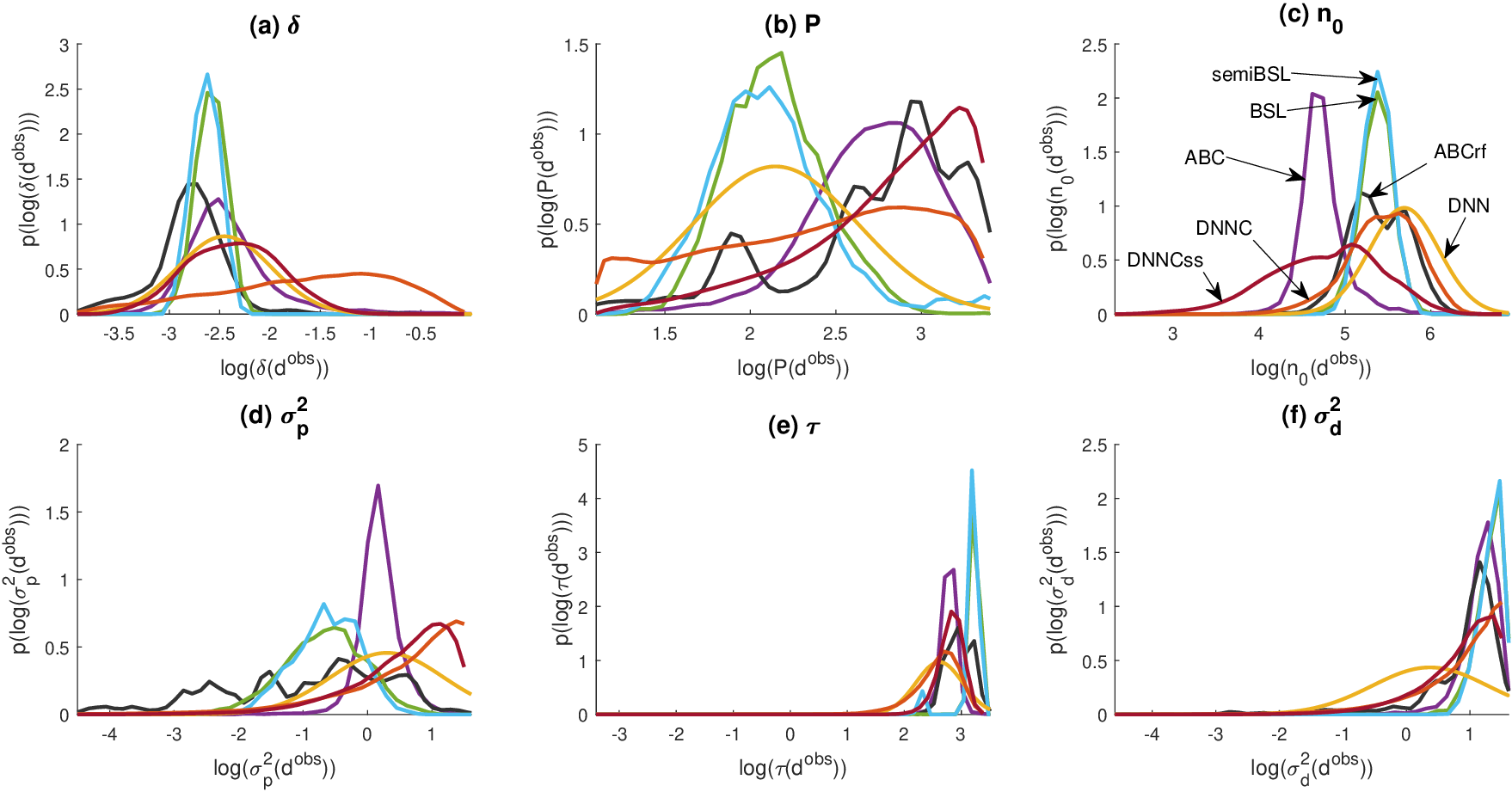}
	\end{center}
	Shown are the marginal posteriors (averages) of log-parameters for the observed blowfly  data  (80\% of the series) for the six methods: BSL (green line), semiBSL (blue line), ABC (purple line), ABCrf (gray line), DNN (yellow line), DNNC (red line), DNNCss (dark red line).
\end{sidewaysfigure}

\begin{sidewaysfigure}[p]
	\caption{Predictive marginal log-posteriors for the observed test data from the voles model.}
	\label{fig:marg:CNN:obs:voles}
	\begin{center}
		\includegraphics[width=0.99\textwidth,angle=0]{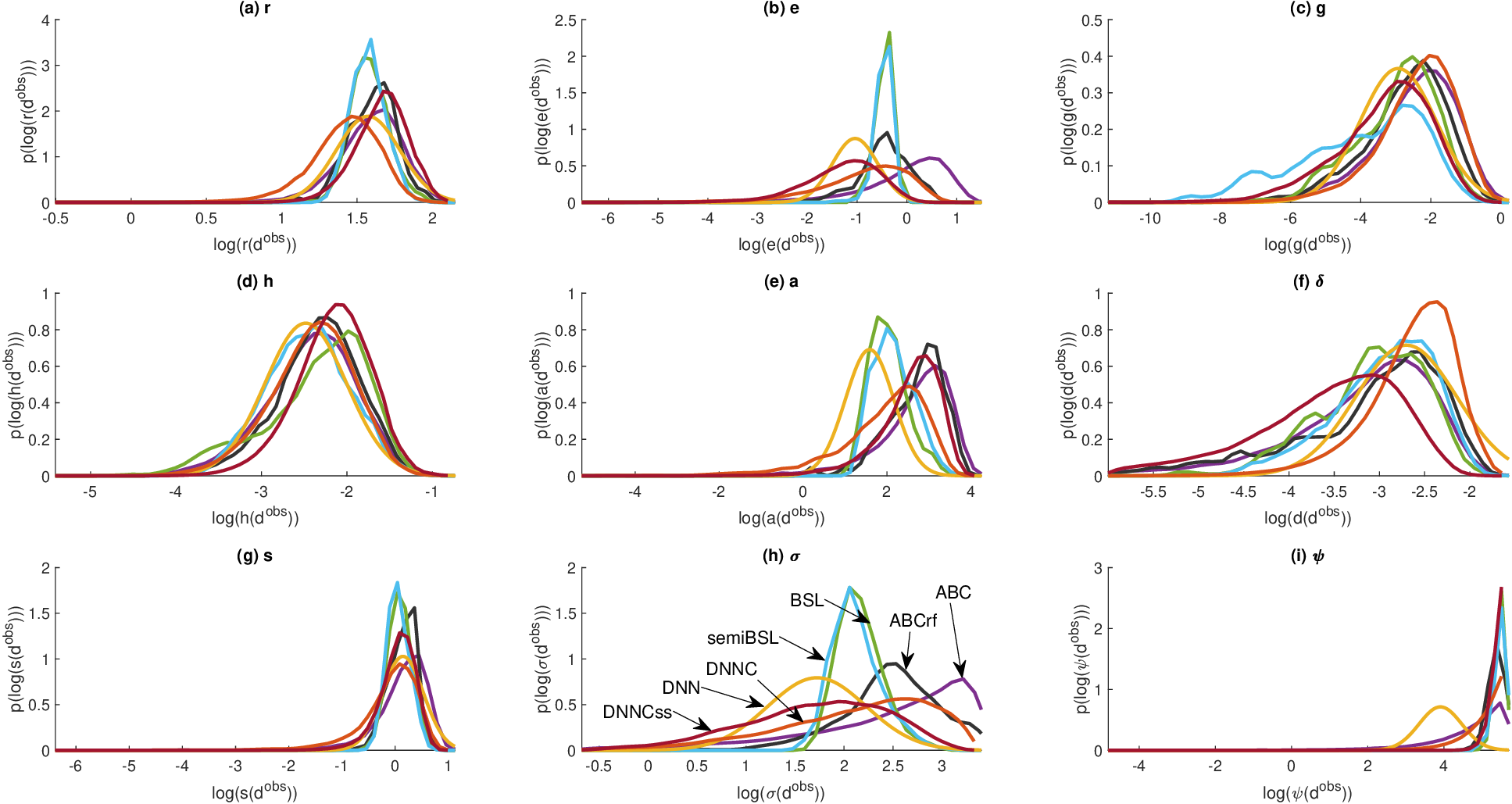}
	\end{center}
	Shown are the  marginal posteriors (averages) of log-parameters for the observed blowfly  data  (80\% of the series) for the six methods: BSL (green line), semiBSL (blue line),  ABC (purple line), ABCrf (gray line), DNN (yellow line), DNNC (red line), DNNCss (dark red line).
\end{sidewaysfigure}

Comparison of the performance of 
different methods when applied to the observed data, is undertaken using data splitting.
Here, the first 80\% of the time series is used as training data, while the last 20\% 
is test data used to assess out-of-sample predictive performance. 
Joint posterior predictive inference is not considered, because 
our own method as well as the ABCrf approach estimates only marginal posterior
distributions. 
Figures~\ref{fig:marg:CNN:obs} and~\ref{fig:marg:CNN:obs:voles} 
show the estimated marginal posterior densities 
of the parameters $\rhovec$ for all competing methods. The figures show that
inferences can differ substantially 
between different methods for some of the parameters, particularly for the blowfly data. 
To examine whether the different
fits are reasonable, our predictive comparison of the different methods using data splitting uses a composite scoring rule \citep{Dawid2014} based on plug-in predictive densities 
using posterior mean point estimates for different methods.  
Consider a sequence of pairwise marginal predictive distributions $\lbrace\hat p(d_{t},d_{t+1}|\hat\rhovec)\rbrace$, with $\hat\rhovec$ being the posterior mean parameters and $(t,t+1)$ consecutive time points in the test part of the series.
With point estimates $\hat{\rhovec}$ obtained from
training data $d_1,\dots, d_K$, and test data consisting of $d_{K+1},\dots, d_T$, out of sample predictive performance is measured by the composite logarithmic score (CLS) and composite energy score (CES) defined as
$$\text{CLS}=1/(T-K-1)\sum_{t=K+1}^{T-1} \log \hat{p}(d_t,d_{t+1}|\hat{\rhovec})\;\;\;\;\text{ and }\;\;\;\;\text{CES}=1/(T-K-1)\sum_{t=K+1}^{T-1} S_t(d_t,d_{t+1}),$$
where $S_t(\cdot,\cdot)$ is the multivariate energy score for the pair $(d_t,d_{t+1})$ \citep{Gneiting2008}. 
The multivariate energy score is the natural generalization
of the continuous ranked probability score to a multivariate setting. 
The composite scores CLS and CES are easier to compute than alternative scoring rules
that involve looking at a full joint distribution for $(d_{K+1},\dots, d_T)^\top$.
The bivariate predictive densities are estimated using kernel estimates based on Monte Carlo simulation from the model. 
We report the scores (with standard errors) in Table~\ref{tab:cnn:scores}, where we compute the negative values 
so that higher values are better. 

\begin{table}
\renewcommand\arraystretch{1.2}
\begin{center}
 \caption{Composite logarithmic scores and negative energy scores for the blowfly and voles data sets.}\label{tab:cnn:scores}
\begin{tabular}{cccccccc}
  \hline\hline
Score  & BSL & semiBSL  & ABC & ABCrf & DNN & DNNC & DNNCss\\\hline
&\multicolumn{7}{l}{{\em Blowfly Data}}\\ \cline{2-8}
	\multirow{ 1}{*}{Log. Score}	& \textbf{  -16.49} & -16.52 & -16.89 & -16.65 & -16.88 & -16.78 & -17.05  \\
	\multirow{ 1}{*}{Neg. Energy} & -1934.34 & -1936.63 & -2380.18 &\textbf{ -1826.64}  & -1979.15  & -1956.89 & -2245.12\\
 &\multicolumn{7}{l}{{\em Voles Data}}\\ \cline{2-8}
  \multirow{ 1}{*}{Log. Score}  &  -11.09 & -11.17 & \textbf{-10.46} & -10.90  & -14.56 & -10.48 & -12.36 \\
\multirow{ 1}{*}{Neg. Energy} &  -48.62 & -49.54 & -42.62 & -44.05 & -61.13 &\textbf{ -37.65} & -58.17 \\
  \hline\hline
\end{tabular}
\end{center}
The columns show the composite scores for the six methods for both
data sets (top: Blowfly; bottom: Voles).
Higher values correspond to more accurate predictive results with the highest values for each case in bold.
\end{table}

\paragraph*{Results}
We make five observations on the simulations. First, our DNNC outperforms all benchmark models, with smallest simulation MSE values, and with coverage rates closest to the nominal 95\% level.
ABCrf was also a strong performer for the voles data, although with
slightly conservative uncertainty assessments. 
Second, DNNC performs better than DNNCss, so that the use of the convolutional network rather than 
user-selected summary statistics makes a difference to the performance of our method.
Third, ABC, DNNC and ABCrf calibrate well marginally, while DNN is poorly 
calibrated in general.
Fourth, simulation MSE values and coverage rates for the ABC and ABCrf are
very similar for the blowfly example. 
Finally, in the comparison of different methods based on the composite scoring rules, DNNC is fourth best for the blowfly data and the best for the voles data for the composite energy score.  
For the composite logarithmic score, DNNC is fourth best for the blowfly data and second best for the voles data.  

\setlength{\abovedisplayskip}{0.1cm}
\setlength{\belowdisplayskip}{0.1cm}
\section{Discussion}\label{sec:discuss}
In this work we have contributed to the growing literature on uncertainty quantification for deep neural network regression models, exploring a marginal calibration
approach using the implicit copula of a deep neural network. Our approach is complementary to existing post-processing adjustments of neural network regression approaches
which attempt to achieve probability calibration.  We have focused particularly on applications to likelihood-free inference for times series models using convolutional networks
to avoid the need for hand-crafted summary statistics. For these
applications the marginal calibration property has a strong motivation as imposing a consistency
requirement on regression approximations to the posterior density that holds for the true posterior density.  

We have only been concerned in this work with regressions for a scalar response.  In the case of our motivating likelihood-free applications, often posterior densities for scalar
functions of the parameter are all that are required for scientific inferences, but the joint posterior may be required for some
interpretive purposes as well as full posterior predictive inference.  It would
be possible to extend the approach considered here to a multivariate response.  
There are at least three ways this could be done.  First, by incorporating some of the simulated parameter values into the features 
in the regression, it would be possible to estimate full conditional posterior densities for individual parameters.  Then
these estimated full conditionals could be used in a likelihood-free Gibbs sampling scheme
\citep{clarte+rrs19,rodrigues+ns20}.  
A related approach involves ordering the parameters, considering a corresponding sequential decomposition of the posterior distribution, and then estimating the univariate conditional distributions in the decomposition using regression. 
A third approach would describe the dependence between components of the response using a copula.  
It would also be interesting in future work to apply the marginal calibration principle
to other regression-based likelihood-free approximations that have been suggested in the literature.

\noindent{\large {\bf Acknowledgments}}\\
The authors would like to thank the Associate Editor and two Referees for inspiring comments that helped to improve upon the first version of this paper. Nadja Klein acknowledges
support through the Emmy Noether grant KL 3037/1-1 of the German research
foundation (DFG). David Nott was supported by a Singapore Ministry of Education
Academic Research Fund Tier 1 grant (R-155-000-189-114). We thank Matteo Fasiolo for sharing his experience on the prior choices, the data and simulators for both likelihood-free data examples.

\appendix
\section{}\label{appA}
This appendix gives some additional details on the implementation
of some of the benchmark methods employed in Section~\ref{sec:LFI}.
\begin{itemize}
\item \underline{DNNCss}: In the FNN, each hidden layer was of size 64, with ReLU/linear activations and dropout rate of 0.5.
\item \underline{ABC}:
We found that using the maximal possible size of 30 for the number of nodes in the hidden layer of the DNN worked best. The $L2$ regularization parameter for the weights in the neural network was chosen by cross-validation, and with a tolerance of one
that corresponds to no rejection step in the ABC procedure, which is recommended for high-dimensional settings. Bounded parameters were transformed to the real line using logistic transformations.
	\item \underline{ABCrf}: For tuning the ABCrf method, we used $500$ trees and tuned several other algorithmic parameters (\texttt{mtry}, the number of randomly chosen features to consider for splits in the trees, \texttt{sample.fraction} the sample fraction of points used in constructing the trees and \texttt{min.node.size}, the node size at which splitting stops) using the \texttt{tuneRanger} package~\citep{tuneRanger}; see the Web Appendix for the optimal hyper-parameter settings obtained. 
	\citet{raynal+mprre18} emphasize increasing training sample size to control variability and monitoring out-of-bag error.  However, in a simulation study it is not possible to adaptively increase the training sample size based on one of the methods.  The default random forests 
	implementation gives poor results with the sample size used here, with overfitting  evident in both examples, but 
	after tuning with the {\tt tuneRanger} package performance was greatly improved.
	\item\underline{BSL} and \underline{semiBSL}: Both are run with $10\mbox{e}^4$ steps and a diagonal covariance for the proposal until convergence. We then take the covariance of these samples as a proposal for the final run.  
\end{itemize}
Both the BSL and semiBSL can experience numerical difficulties if the MCMC starting value is in the posterior tail, so we initialized these using the posterior mean parameters of ABC. 
\renewcommand{\baselinestretch}{-1.0}
\onehalfspacing
\bibliography{references}
\end{document}